\documentclass[twocolumn,twocolappendix]{aastex631}

\usepackage{graphicx} 
\usepackage{bm}       
\usepackage{CJK}

\newcommand{\p}{\bm{p}}

\newcommand{\cgpd}[1]{\Phi\left(#1|M,z\right)}
\newcommand{\Msun}{M_{\odot}}
\newcommand{\nhalo}{n_{\rm halo}}

\newcommand{\mhh}{m_{{\rm H}_2}}
\newcommand{\hh}{{\rm H}_2}
\newcommand{\rgal}{r_{\rm gal}}
\newcommand{\sfr}{{\rm SFR}}
\newcommand{\ssfr}{{\rm sSFR}}

\newcommand{\cov}{{\rm Cov}}

\newcommand{\gkai}[1]{\begin{CJK*}{UTF8}{gkai}\raisebox{.1em}{(}#1\raisebox{.1em}{)}\end{CJK*}}

\shorttitle{Characterizing CGPD using GMM}
\shortauthors{Zhang et al.}

\begin{document}

\title{Characterizing the Conditional Galaxy Property Distribution using Gaussian Mixture Models}

\correspondingauthor{Yucheng Zhang}
\email{yucheng.zhang@nyu.edu}

\author[0000-0002-9300-2632]{Yucheng Zhang \gkai{张宇澄}}
\affiliation{Department of Mathematics and Theory, Peng Cheng Laboratory, Shenzhen, Guangdong 518066, China}
\affiliation{Center for Cosmology and Particle Physics, Department of Physics, New York University, 726 Broadway, New York, NY 10003, USA}

\author[0000-0002-2091-8738]{Anthony R. Pullen}
\affiliation{Center for Cosmology and Particle Physics, Department of Physics, New York University, 726 Broadway, New York, NY 10003, USA}
\affiliation{Center for Computational Astrophysics, Flatiron Institute, New York, NY 10010, USA}

\author[0000-0002-6748-6821]{Rachel S. Somerville}
\affiliation{Center for Computational Astrophysics, Flatiron Institute, New York, NY 10010, USA}

\author[0000-0001-8382-5275]{Patrick C. Breysse}
\affiliation{Center for Cosmology and Particle Physics, Department of Physics, New York University, 726 Broadway, New York, NY 10003, USA}

\author[0000-0002-1975-4449]{John C. Forbes}
\affiliation{Center for Computational Astrophysics, Flatiron Institute, New York, NY 10010, USA}

\author[0000-0002-0782-9116]{Shengqi Yang \gkai{杨晟祺}}
\affiliation{Carnegie Observatories, 813 Santa Barbara Street, Pasadena, CA 91101, USA}

\author[0000-0002-0701-1410]{Yin Li \gkai{李寅}}
\affiliation{Department of Mathematics and Theory, Peng Cheng Laboratory, Shenzhen, Guangdong 518066, China}
\affiliation{Center for Computational Astrophysics, Flatiron Institute, New York, NY 10010, USA}
\affiliation{Center for Computational Mathematics, Flatiron Institute, New York, NY 10010, USA}

\author[0000-0002-4617-9320]{Abhishek S. Maniyar}
\affiliation{Center for Cosmology and Particle Physics, Department of Physics, New York University, 726 Broadway, New York, NY 10003, USA}

\begin{abstract}

Line-intensity mapping (LIM) is a promising technique to constrain the global distribution of galaxy properties.
To combine LIM experiments probing different tracers with traditional galaxy surveys and fully exploit the scientific potential of these observations, it is necessary to have a physically motivated modeling framework. As part of developing such a framework, in this work we introduce and model the conditional galaxy property distribution (CGPD), i.e. the distribution of galaxy properties conditioned on the host halo mass and redshift.
We consider five galaxy properties, including the galaxy stellar mass, molecular gas mass, galaxy radius, gas phase metallicity and star formation rate (SFR), which are important for predicting the emission lines of interest.
The CGPD represents the full distribution of galaxies in the five dimensional property space;  many important galaxy distribution functions and scaling relations, such as the stellar mass function and SFR main sequence, can be derived from integrating and projecting it.
We utilize two different kinds of cosmological galaxy simulations, a semi-analytic model and the IllustrisTNG hydrodynamic simulation, to characterize the CGPD and explore how well it can be represented using a Gaussian mixture model (GMM).
We find that with just a few ($\sim 3$) Gaussian components, a GMM can describe the CGPD of the simulated galaxies to high accuracy for both simulations. The CGPD can be mapped to LIM or other observables by constructing the appropriate relationship between galaxy properties and the relevant observable tracers. 
\end{abstract}


\section{Introduction}

Understanding the physical processes that govern the formation and evolution of galaxies over cosmic time is one of the most fundamental and important problems in astrophysics and cosmology.
Numerous astronomical surveys across a wide range of wavelengths have been carried out to observe galaxies, from nearby targets to those probing the earliest epochs~\citep[e.g.][]{York2000,Koekemoer2011}.
However, galaxy surveys are limited by the trade-off between area and depth since deeper and fainter targets require more time to achieve a detectable signal-to-noise ratio.
Instead of resolving and observing individual targets, a relatively recent technique called line-intensity mapping (LIM)~\citep[e.g.][]{Kovetz2017} has been proposed to measure the integrated emission lines from all sources, bright and faint.
Similar to traditional galaxy surveys, LIM not only provides a promising tracer of the large-scale structure (LSS) of the total matter distribution, but also contains a wealth of information about astrophysical properties of galaxies~\citep{Schaan2021a}.

One of the potentially powerful aspects of LIM is that there are now numerous experiments planned or online that probe different lines, which trace different gas phases, at different redshifts \citep[see][for a recent summary]{bernal:2022}. For example, some of the most commonly considered lines include CO, which is a tracer of dense molecular gas in the interstellar medium (ISM) of galaxies, atomic and ionic fine structure lines such as [CII], which is primarily emitted by photo-dissociation regions in the outer layers of molecular clouds, 21-cm emission from atomic gas in the ISM and intergalactic medium (IGM), and lines arising from ionized gas such as Lyman-$\alpha$ and H-$\alpha$. These are the brightest and most easily detected lines, but there are many others that can provide interesting information. It is clear that combining multi-tracer LIM surveys with traditional galaxy surveys will be the key to extracting the maximum amount of science from these experiments. 

In order to do this, it is essential to have a comprehensive modeling framework. The most commonly used framework is halo-based phenomenological models \citep[see][for a recent review]{bernal:2022}. In this approach, scaling relations between halo mass and redshift and ``intrinsic'' physical properties such as stellar mass and star formation rate are derived using abundance matching. In abundance matching, observational estimates of stellar mass functions and SFR are used along with predictions for the number density of dark matter halos as a function of mass to derive constraints on the mapping between these quantities \citep[see][for a review]{wechsler-tinker:2018}. Then, scaling relations between e.g. SFR and a line luminosity of interest are derived using observations of individually detected sources~\citep[e.g.][]{Silva2015,Li2016,Padmanabhan2018,Padmanabhan2019,Sun2019}. This approach has the advantage of being highly computationally efficient, and of being tied to observations.

However, there are a number of drawbacks and limitations to this approach. First, these empirical fitting functions usually lack physical motivation, so they are able to provide only limited insights about the physical processes underlying galaxy formation. Second, LIM detections are sensitive to galaxies that are too faint to detect individually, so the scaling relations have to be extrapolated beyond the observational range. Third, these empirical approaches are not able to self-consistently capture the covariances between different tracers, or between the intrinsic properties (stellar mass, SFR) and various emission lines. It is important to keep in mind that the estimates of stellar mass and SFR that are used in abundance matching models are not directly observable quantities, but are derived from UV-optical-NIR observations and carry significant uncertainties. The prior used to obtain these estimates can have a strong impact on the results \citep{Leja2022}.  It is critical to be able to properly fold in the prior, covariances, and uncertainties on these quantities, which lie at the foundation of abundance matching approaches. 

Numerical cosmological simulations that incorporate baryonic  astrophysical processes are a potentially powerful tool for exploring the physics of galaxy formation and evolution, as well as the connection with observable line emissions.
The most fundamental approach is based on explicitly simulating the hydrodynamics and thermodynamics of baryonic material evolving within a $\Lambda$CDM cosmological framework for structure formation \citep{somerville-dave,naab2017}. These numerical hydrodynamic simulations provide direct predictions for the distribution and state of nearly all of the relevant components of galaxies and the IGM (stars, gas at different densities and temperatures, various chemical elements). They can be coupled with modeling tools to provide predictions of many of the emission lines of interest for LIM experiments \citep{Leung,Olsen}. However, due to its computational expense, this technique suffers from a fairly severe trade-off between volume and resolution. Simulations that resolve the multi-phase ISM can typically be run for only a few galaxies \citep[e.g.][]{pallottini}. Simulations of cosmological volumes must adopt ``sub-grid'' recipes to treat most of the critical processes of galaxy formation, including star formation, stellar feedback, and black hole formation and feedback. And even the largest computationally feasible cosmological hydrodynamic simulations fall short of the volumes that are already routinely probed by galaxy and LIM surveys.  For example, MillenniumTNG has a volume of (500 Mpc/$h$)$^3$, compared to (1760 Mpc/$h$)$^3$ for the Baryon Oscillations Spectroscopic Survey of $z<0.7$ luminous red galaxies alone \citep{2017MNRAS.470.2617A}. The currently adopted sub-grid models are highly uncertain, and approaches adopted by different groups can lead to dramatically different predictions for gas properties. It is infeasible to thoroughly explore the parameter space of the sub-grid models for significant volumes using a direct numerical approach. 

An alternative approach to physics-based models is semi-analytic models (SAM), which combine simplified treatments of physical processes with cosmological halo merger histories. These models produce similar predictions to those of numerical hydrodynamic simulations for global galaxy properties, and include a similar suite of physical processes, but at a fraction of the computational cost. However, unlike numerical hydrodynamics simulations they do not provide detailed information about the internal structure of galaxies or the spatial distribution of gas within or around them. \citet{Popping2014} and \citet{popping2019} coupled semi-analytic models of galaxy formation with the {\sc DESPOTIC} code \citep{despotic}, which computes line luminosities for optically thick clouds using a one-zone model, to make predictions for a broad suite of emission lines, including multiple transitions in the CO ladder, [CI] and [CII]. They validated these models by comparing the predictions with an extensive suite of observations of individually detected galaxies across a broad range of redshifts. \citet{Yang2021} used lightcones populated with this approach to create mock maps (including interlopers, cosmic infrared background, and Galactic foreground) for exemplar LIM experiments in CO and [CI]. This work showed that SAMs coupled with the appropriate line emission models in post-processing are a promising alternative tool for LIM forecasts and interpretation. 

Ideally, we would like to be able to use Bayesian inference to constrain the posteriors of both cosmological and astrophysical parameters from a combined suite of multi-tracer LIM and galaxy survey observables. When using a forward-modeling or simulation-based inference approach on a large population of galaxies or a cosmological volume \citep{2009MNRAS.396..535H,2013MNRAS.431.3373H,2011MNRAS.416.1949L,2012MNRAS.421.1779L, 2014MNRAS.443.1252L,2019MNRAS.487.3581F}, depending on the sampling approach adopted, this requires thousands to tens of thousands of executions of the simulation. Most currently available SAM codes are not fast enough for this to be practical, particularly if large volumes need to be simulated. 

An intermediate approach is to develop models that link halo properties to galaxy intrinsic and observable properties using physics-based simulations, such as hydrodynamic simulations and SAMs, instead of empirical methods such as abundance matching. For example, \citet{yang2022} used the SAM-based predictions described above to derive and present fitting functions between halo mass and all major predicted lines over a broad range of redshifts, and showed that when these fitting functions were combined with dark matter halo properties as in the standard empirical halo model approach, agreement of 5-10\% was achieved for common LIM summary statistics such as the power spectrum and intensity. 

\begin{figure*}
    \centering
    \includegraphics[width=\textwidth]
    {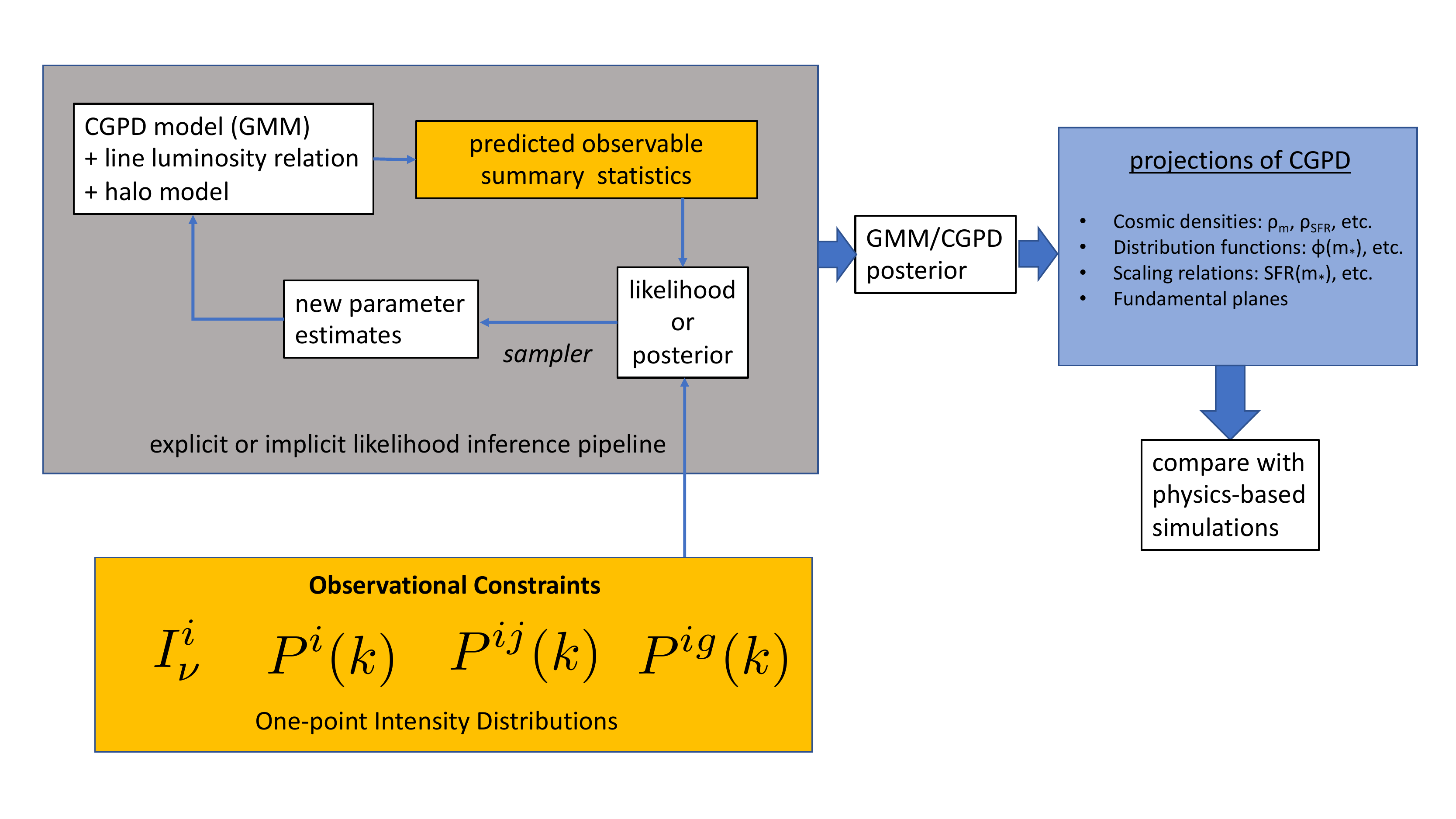}
    \caption{Illustration of how a parametric model of the Conditional Galaxy Property Distribution can be used to obtain insights about the evolution of physical quantities of interest. First, a general class of models for representing the CGPD (here GMM) is motivated using physics-based models of galaxy formation. The models must be flexible but not overly complex. The CGPD is combined with a model that predicts the observed line luminosity as a function of the galaxy properties (the Line Luminosity Relation, LLR), to predict observable summary statistics, such as the power spectrum, one-point distribution, or integrated line intensity. Observational estimates of these quantities are then used to estimate the likelihood (if using a ``traditional" explicit likelihood inference approach) or to directly estimate the posterior, if using implicit likelihood inference. Using a sampler, a new set of parameter values is guessed, and the model is run again until convergence is reached. Now the ``best fit'' CGPD can be projected and integrated as described in Section~\ref{sec:cgpd}, yielding the posterior and confidence intervals on quantities of great interest for galaxy evolution, as summarized in the rightmost box.  These estimates can be compared to traditional estimates from galaxy surveys, where physical parameters are estimated through SED fitting, and with the predictions of physics-based simulations. 
    }
    \label{fig:inference_diagram}
\end{figure*}

In this work, we introduce a new approach that we believe can have very general applicability to forecasting and interpreting a wide variety of galaxy and LIM survey data. 
We introduce the conditional galaxy property distribution (CGPD) $\cgpd{\p}$, which gives the number density of galaxies in the property space $\p$, conditioned on the halo mass $M$ and redshift $z$. Whereas traditional halo models are typically mappings from a scalar quantity (halo mass) to another scalar (e.g. stellar mass or SFR), we can think of this as an extension of the halo model to map to a higher dimensional suite of galaxy properties. 
In this work, $\p$ has dimension five and includes the galaxy stellar mass, molecular gas mass, galaxy radius, gas phase metallicity and star formation rate (SFR), which are important for predicting the line luminosities of interest (here, we have in mind mainly CO and [CII]). However, in principle $\p$ could contain any galaxy property of interest and have an arbitrary number of dimensions. With this set of galaxy properties, CGPD also encapsulates many popular and important galaxy distribution functions and scaling relations such as the stellar mass function and SFR main sequence. 

In this paper, we explore the form and effective dimensionality of the CGPD for this chosen set of key galaxy properties, using two different physics based cosmological simulations of galaxy formation: the ``Santa Cruz'' semi-analytic model \citep{Somerville1999,Somerville2008,Somerville2012,Porter2014,Popping2014,Somerville2015a}, and the IllustrisTNG hydrodynamic simulation suite \citep{TNG-1,TNG-2,TNG-3,TNG-4,TNG-5}. A key point is that it is well known that galaxy properties are highly covariant, giving rise to the many familiar galaxy scaling relations such as the stellar mass vs. SFR relation (SF sequence), mass metallicity relation, etc. Exploring higher dimensional manifolds within galaxy property space (as hinted at by, for example, ``fundamental plane'' type relations such as the fundamental mass metallicity relation) will in itself provide interesting insights into galaxy formation, as the physics that shapes galaxy properties is encoded in these relationships. Moreover, it will help us to develop tools to characterize and model these high-dimensional relationships in order to fully capture the covariances between galaxy properties. In this work, we adopt a Gaussian mixture model (GMM) based approach as a generative density estimation model to describe the CGPD.

Combined with a model that can predict emission line luminosities from these galaxy properties, this parametric model could potentially be used in a statistical inference pipeline, either using ``traditional" methods such as Markov Chain Monte Carlo (MCMC) sampling combined with an assumed form for the likelihood, or more modern implicit likelihood inference methods \citep{Alsing2018,2020PNAS..11730055C}. With the formulation presented here, the posterior of the parameters of the GMM (and hence the CGPD) can be constrained based on a chosen set of LIM observables. The resulting CGPD can then be projected and integrated to obtain estimates for galaxy population summary statistics of common interest, such as the SFR or cold gas distribution functions or cosmic SFR or cold gas density, at a series of redshifts. If desired, additional observational constraints from galaxy surveys, such as the stellar mass function, can simultaneously be used as constraints in the inference procedure. In Fig.~\ref{fig:inference_diagram}, we provide a graphical representation of how the CGPD model that we develop in this work fits into this broader framework, which we plan to build up in future works.

The paper is organized as follows.
In Section~\ref{sec:cgpd}, we introduce the CGPD and how it is related to more familiar summary statistics that are commonly used to characterize galaxy populations. 
We introduce the simulated galaxy catalogs with physical properties of interest in Section~\ref{sec:simulations}.
Based on these simulations, we then present the application of the GMM in characterizing the CGPD in Section~\ref{sec:cgpd_model}.
We discuss our results in Section~\ref{sec:discussion}, and conclude in Section~\ref{sec:conclusions}.
Throughout this work, we adopt a flat $\Lambda$CDM fiducial cosmology with \textit{Planck} 2015 parameters: $\Omega_m=0.3089$, $\Omega_\Lambda=0.6911$, $\Omega_b=0.0486$, $h=0.6774$, $\sigma_8=0.8159$, $n_s=0.9667$~\citep{Planck2015-cosmo}, which are consistent with the cosmology assumed in the simulations.

\section{Conditional Galaxy Property Distribution} \label{sec:cgpd}

For a halo of mass $M$ at redshift $z$, the CGPD $\cgpd{\p}$ is defined as the number density distribution of galaxies in the property space $\p$.
Integrating $\cgpd{\p}$ over $\p$ yields the expected number of galaxies for such a halo.
For simplicity, in this work we write the CGPD as a function of $\p$, although in practice, we perform the fitting of the CGPD in $\tilde{\p} = \log\p$ space~\footnote{In this work, we denote a base 10 logarithm with $\log$ and natural logarithm with $\ln$.}, i.e. $\tilde{\Phi}(\tilde{\p}|M,z)$.
The CGPD may be represented in either of the two spaces for convenience.
The transformation of the CGPD between these two spaces can be straightforwardly given by
\begin{equation}
    \cgpd{\p} = \tilde{\Phi}\left(\tilde{\p}(\p)|M,z\right)\left| \frac{\partial\,\tilde{\p}(\p)}{\partial\,\p} \right| \,,
\end{equation}
where $\left| \partial\,\tilde{\p}(\p) / \partial\,\p \right|$ is the determinant of the Jacobian matrix.
Multiplying the differential volume in $\p$ space on both sides gives
\begin{equation}
    \cgpd{\p} \prod_i\,dp_i = \tilde{\Phi}\left(\tilde{\p}|M,z\right) \prod_i\,d\tilde{p}_i \,,
\end{equation}
which is more convenient for the purpose of integration.
The galaxy properties we consider are summarized in Table~\ref{tab:properties}.
\begin{table}
    \centering
    \caption{The 5-D galaxy properties $\p$ for the CGPD $\cgpd{\p}$ considered in this work, which are selected based on their relevance to the emission lines of interest.}
    \label{tab:properties}
    \begin{tabular}{cll}
        \hline\hline
        Property & Unit & Description \\
        \hline
        $m_*$     & $[\Msun]$ & galaxy stellar mass \\
        $\mhh$ & $[\Msun]$ & molecular gas mass \\
        $\rgal$   &  [kpc] & galaxy radius \\
        $Z_g$     & $[Z_{\odot}]$ & gas phase metallicity \\
        $\sfr$    & $[\Msun/{\rm yr}]$ & star formation rate \\
        \hline
    \end{tabular}
\end{table}
These are the properties that are the most important for the line luminosities of interest.
Note that $\cgpd{\p}$ is related to the conditional luminosity function (CLF) $\Phi(\bm{L}|M,z)$ introduced in~\cite{Schaan2021b} by the line luminosity relation (LLR) $L_i(\p,z)$, which gives the luminosity of line $i$ for a galaxy with properties $\p$ at redshift $z$.
The LLR is assumed to be independent of the host halo mass $M$.
The modeling of the LLR using SAM simulated galaxies coupled with spectral synthesis models is being developed in a companion work~\citep{Breysse2022}.

The CGPD must be coupled with the dark matter halo model in order to connect the cosmological dark matter field with observable galaxies.
For example, we obtain the global distribution function for each element of the property vector $\p$ by performing the integral:
\begin{equation} \label{eq:phi}
    \phi(\p|z) = \int dM\, \nhalo(M, z) \cgpd{\p} \,,
\end{equation}
where $\nhalo(M, z)$ is the halo mass function (HMF) at redshift $z$.
In this work, we use the HMF from~\cite{Tinker2008} with the implementation in the \texttt{hmf}~\footnote{\url{https://github.com/halomod/hmf}} code~\citep{Murray2013}.
This general galaxy distribution function $\phi(\p|z)$ gives the comoving volume number density of galaxies in the property space $\p$ at redshift $z$.
Integrating $\phi(\p|z)$ over $\p$ yields the total number of galaxies per unit comoving volume.
As shown below, by integrating over certain dimensions, $\phi(\p|z)$ encapsulates many observables of wide interest in galaxy and LIM surveys, such as the cosmic densities of galaxy properties, the galaxy abundance as a function of a certain property, and the scaling relations between two (or more) galaxy properties.

\subsection{Cosmic densities of galaxy properties}

The cosmic number density of galaxies per unit comoving volume is given by
\begin{equation} \label{eq:gal_vol_den}
    n_g(z) = \int d\p\, \phi(\p|z) \,.
\end{equation}
Similarly, the cosmic density of a galaxy property $p_i$ in $\p$ is
\begin{equation} \label{eq:prop_cosmic_den}
    \rho_{p_i}(z) = \int d\p\, \phi(\p|z)p_i \,.
\end{equation}
These integrated quantities are important observable functions describing the overall evolution of galaxy properties over cosmic time~\citep[e.g.][and references therein]{Madau2014}, including
the cosmic stellar mass density $\rho_{m_*}(z)$,
cosmic molecular gas mass density $\rho_{m_{\hh}}(z)$,
and cosmic SFR density $\rho_\sfr(z)$.

\subsection{Galaxy distribution functions and scaling relations}

Besides the volume density of the galaxy properties in Eq.~\ref{eq:prop_cosmic_den}, the abundances of galaxies as functions of certain physical properties have also been widely studied in galaxy surveys.

As a function of a single property $p_i$, the galaxy distribution $\phi(p_i|z)$ is given by integrating $\phi(\p|z)$ over the other $\dim(\p)-1$ dimensions,
\begin{equation} \label{eq:phi_1}
    \phi(p_i|z) = \int \phi(\p|z) \, \prod_{j\neq i} dp_j \,.
\end{equation}
These functions of wide interest include
the galaxy stellar mass function $\phi(m_*)$~\citep[e.g.][]{Baldry2008},
the molecular gas mass function $\phi(\mhh)$~\citep[e.g.][]{Diemer2018},
and the star formation rate function $\phi(\sfr)$~\citep[e.g.][]{Smit2012}, etc.

Another important observable for galaxy surveys is the correlation between certain sets of galaxy properties, some of which can be even described with simple scaling relations.
At a given redshift, integrating $\phi(\p|z)$ over $\dim(\p)-2$ dimensions of $\p$ yields
\begin{equation} \label{eq:phi_2}
    \phi(p_i, p_j|z) = \int \phi(\p|z) \, \prod_{k\neq i,j} dp_k \,,
\end{equation}
the distribution of galaxies in the space of the two unprojected-over properties.
A traditional scaling relation corresponds to fitting this distribution with an empirical function $p_i(p_j)$.
These scaling relations of wide interest include
the star formation main sequence $\sfr(m_*)$~\citep[e.g.][]{Noeske2007}, 
the mass-metallicity relation $Z_g(m_*)$~\citep[e.g.][]{Tremonti2004},
the size-mass relation $\rgal(m_*)$~\citep[e.g.][]{Lange2015,Mowla2019},
and the stellar mass vs. gas mass relation $m_* - \mhh$~\citep[e.g.][]{Calette2018}.
Besides the main trend or correlation given by the scaling relations, there is also dispersion around the median values.
These dispersions $\sigma_{p_i}(p_j)$ also contain a wealth of physical information~\citep[e.g.][]{Forbes2014}, and the corresponding conditional distributions like $\phi(p_i|p_j)$ have also been investigated~\citep[e.g.][]{Leja2022}.
In our case, $\phi(p_i, p_j)$ includes all the information, from which the scaling relations and the conditional distribution can also be derived.
The conditional probability distribution $f(p_i|p_j)$ is given by
\begin{equation} \label{eq:cond_ij}
    f(p_i|p_j) = \frac{\phi(p_i,p_j)}{\int dp_i\, \phi(p_i,p_j)} \,,
\end{equation}
where $f(p_i|p_j)$ differs from $\phi(p_i|p_j)$ by a factor of the galaxy number density.
Then the function $p_i(p_j)$ can be modelled as
\begin{equation} \label{eq:scaling_ij}
    \left< p_i(p_j)\right> = \int dp_i\, p_i f(p_i|p_j) \,.
\end{equation}
Similarly, the dispersion follows,
\begin{equation} \label{eq:scatter_ij}
    \sigma^2_{p_i}(p_j) = \left<p^2_i(p_j)\right> - \left<p_i(p_j)\right>^2 \,.
\end{equation}

Integrating over $\dim(\p)-3$ dimensions yields $\phi(p_i,p_j,p_k|z)$, which leads to ``fundamental plane'' relations $p_i(p_j, p_k)$ such as $m_*(Z_g, \mhh)$ or $m_*(Z_g, \sfr)$~\citep[e.g.][]{Lara-Lopez2010}.
The expressions are almost the same as Eqs.~\ref{eq:cond_ij},~\ref{eq:scaling_ij} and~\ref{eq:scatter_ij}, with $p_j$ simply replaced with $p_j,\,p_k$.

\subsection{Line luminosities and intensities}


Typically, line luminosity functions are 1-D distributions over one line luminosity $L$.
As with the relation between CGPD and CLF mentioned above, the line luminosity function $\phi(L|z)$ can also be connected to $\phi(\p|z)$ by the LLR $L_0(\p, z)$,
\begin{equation}
    \phi(L|z) = \int d\p\, \delta^{\rm D}(L-L_0(\p,z)) \phi(\p|z) \,.
\end{equation}

By combining $\phi(\p|z)$ and LLR, the observed intensity can be derived as
\begin{equation} \label{eq:line_intensity}
    I_\nu(z) = \frac{1}{4\pi\nu_0}\frac{c}{H(z)}\int d\p\, \phi(\p|z) L_0(\p, z) \,,
\end{equation}
where the integral (without the factor) gives the line luminosity density.

\subsection{Galaxy properties per halo}

Besides the expressions based on Eq.~\ref{eq:phi} which integrate over all halos of different masses, we can similarly write down the properties per halo from the CGPD $\cgpd{\p}$.
For example, the average number of galaxies for a halo of a given mass can be written as
\begin{equation} \label{eq:N_gal_halo}
    N_g^h(M, z) = \int d\p\, \cgpd{\p} \,,
\end{equation}
where the superscript $h$ indicates that it is a quantity per halo.
Similarly, the expected galaxy property $p_i$ per halo is given as
\begin{equation} \label{eq:prop_halo}
    P_i^h(M, z) = \int d\p \, \cgpd{\p} p_i \,.
\end{equation}
Note that Eq.~\ref{eq:prop_halo} should be divided by Eq.~\ref{eq:N_gal_halo} for $\rgal$ if the average radius per galaxy is desired instead of the sum of all galaxy radii in the halo.
With LLR $L_i(\p, z)$, the line luminosity for line $i$ per halo can be written as
\begin{equation} \label{eq:line_halo}
    L_i^h(M,z) = \int d\p\,\cgpd{\p}L_i(\p, z) \,.
\end{equation}

\section{Simulations} \label{sec:simulations}

In this section we briefly introduce the physics-based simulations, which will be used to study and motivate the functional form and modeling approach for characterizing the CGPD. These include the ``Santa Cruz" semi-analytic models and hydrodynamic simulations from the IllustrisTNG project.
Note that comparing and discussing the differences between the simulated results from the SAM and TNG is outside of the scope of this work, and a recent discussion can be found in~\cite{Gabrielpillai2021}.
It is important to consider different models in order to test the robustness and flexibility of the GMM in describing the CGPD.

\subsection{Semi-Analytic Model} \label{subsec:sam}

State-of-the-art SAMs have been shown to produce consistent predictions for global galaxy properties and their evolution compared with much more expensive numerical hydrodynamic simulations~\citep{Somerville2015b}.
In this work, we use the SAM developed by the ``Santa Cruz'' group~\citep{Somerville1999,Somerville2008,Somerville2012,Porter2014,Popping2014,Somerville2015a}.
This well-established semi-analytic galaxy formation modeling code has been compared extensively with galaxy observations from $z \sim 0 - 10$ and is overall very successful at matching many observations~\citep[see e.g.][]{Popping2014,Somerville2015a,Yung2019a,Yung2019b,Somerville2021}.

The backbone of the SAM consists of dark matter halo merger trees, which can be extracted from N-body simulations or generated with Monte Carlo realizations based on the extended Press-Schechter (EPS) formalism~\citep{Somerville1999a}.
In this work, in order to make sure that all halo masses in the wide range of interest are well populated for fitting, we use the Monte Carlo EPS method, which enables better sampling of the full range of relevant halo masses at low computational cost.

We generate simulations for one hundred redshifts uniformly sampled in $0 \leq z < 10$, with galaxies at each redshift being simulated independently.
At each redshift, we consider halos with mass in the range $10 \leq \log(M/\Msun) \leq 13$, which has been shown to be the relevant halo mass range probed by LIM measurements~\citep{Yang2021}.
In this logarithmic halo mass range, a thousand halos are generated for each of the one thousand uniformly sampled halo masses, and hence in total this yields one million halos for each redshift.
This large number of halos and well-sampled resolution in halo mass give us more flexibility in binning for constructing the CGPD.
We only consider star-forming galaxies, which are selected with the criterion $\ssfr > 1/(3t_H(z))$, where $\ssfr \equiv \sfr/m_*$ is the specific SFR and $t_H(z)$ is the Hubble time at the galaxy redshift $z$.
We ignore quenched galaxies in this work, as they have a negligible contribution to LIM.
For simplicity and better demonstration, in our main analyses we only consider central galaxies.
We checked that the contributions from satellites for the integrated quantities are less than $10\%$, and we include some further details on including and fitting satellites in Section~\ref{sec:satellites}. 

\subsection{Hydrodynamic simulations} \label{subsec:tng}
IllustrisTNG is a project that includes a series of state-of-the-art cosmological N-body and hydrodynamic simulations~\citep{TNG-1,TNG-2,TNG-3,TNG-4,TNG-5}.
We use the halo and subhalo (i.e. galaxy) catalogs from TNG100-1, which are simulated in a cubic box of size $75\,{\rm Mpc/h}$ with a mass resolution of $\sim 10^6\,\Msun$ for the dark matter and baryonic particles.
The TNG fields that are most comparable to the galaxy properties simulated by the SAM used in this work are summarized in Table~\ref{tab:tng_fields}.
\begin{table}
    \centering
    \caption{Fields in TNG subhalo catalogs that are most comparable to the galaxy properties in Table~\ref{tab:properties}.
    The units have also been transformed correspondingly.}
    \label{tab:tng_fields}
    \begin{tabular}{cll}
        \hline\hline
        Property & TNG field \\
        \hline
        $m_*$     & \texttt{SubhaloMassInRadType (Type=4)} \\
        $\mhh$ & \texttt{m\_h2\_DK11\_vol} \\
        $\rgal$   & \texttt{SubhaloHalfmassRadType (Type=4)} \\
        $Z_g$     & \texttt{SubhaloGasMetallicity} \\
        $\sfr$    & \texttt{SubhaloSFR} \\
        \hline
    \end{tabular}
\end{table}
TNG fields for $m_*$ and $\sfr$ are selected following the discussion in~\cite{Gabrielpillai2021}, where it is also noted that in some cases these quantities are not defined in exactly the same way as in SAM.
Then the field for $\rgal$ is chosen to be consistent with that of $m_*$.
The molecular gas mass $\mhh$ estimates are provided in post-processed supplementary catalogs~\citep{Diemer2018,Diemer2019,Stevens2021} for five redshifts $z=(0,\ 0.5,\ 1,\ 1.5,\ 2)$, which are described in detail in the data specifications of the TNG project~\footnote{\url{https://www.tng-project.org/data/docs/specifications/}}.
For the host halo mass $M$, the \texttt{Group\_M\_TopHat200} field in TNG is used.

\section{Results: Characterizing the CGPD} \label{sec:cgpd_model}

Based on the simulated data described above, we now propose and test analytic models that could potentially properly characterize the CGPD.

There are various parametric and nonparametric methods for describing the distribution of point data, also known as density estimation~\citep{astroMLText}.
Histograms might be the most straightforward nonparametric method, which approximate the function value as constants within discrete bins.
However, the total number of bins increases exponentially with the dimensionality of the data.
Treating each bin as an independent variable makes it infeasible for inference, and also this method does not give any insights into the structure of the distribution.
This remains an issue even for more advanced nonparametric models like Gaussian processes.
Another class of methods assumes a specific parametric form of the underlying probabilistic density function that generates the data points, among which Gaussian Mixture Models are one of the most powerful approaches due to its simplicity and flexibility.
In this section, we present the modeling of the CGPD using GMMs.

\subsection{Gaussian Mixture Model}

The GMM is a probabilistic model defined as the weighted average of a finite number of Gaussian distributions, and the likelihood function reads
\begin{equation} \label{eq:gmm_def}
    L(\bm{x}|\bm{\theta}) = \sum_{j=1}^{N_{\rm gc}} w_j\, \mathcal{N}(\bm{x}|\bm{\mu}_j,\bm{\Sigma}_j) \,,
\end{equation}
where $\mathcal{N}$ denotes a multivariate Gaussian distribution, $\bm{x}$ is the data vector, and $\bm{\theta}$ includes all the model parameters, i.e. the weight $w_j$, the mean vector $\bm{\mu}_j$ and the covariance matrix $\bm{\Sigma}_j$ for each of the $N_{\rm gc}$ Gaussian components.
The weights are constrained by the normalization $\sum_{j=1}^{N_{\rm gc}} w_j = 1$ such that the likelihood function integrates to one.
In unsupervised machine learning, GMM is a powerful method for clustering and classification with each component corresponding to a certain class.
Here, GMM is mainly used for density estimation and we are not attempting to classify galaxies in their property space.
Another practical advantage of GMM in describing multidimensional data is their simplicity in giving marginal and conditional distributions, which remain Gaussian.
Since each component is a multivariate Gaussian, marginalizing (integrating) over any subset of dimensions can be done by simply removing the marginalized dimensions from the mean vector and covariance matrix.
Similarly, the conditional distributions for a multivariate Gaussian can also be written analytically.
For example, for a bivariate case,
\begin{equation} \label{eq:gaussian_cond}
    f(X_1|X_2=x_2) \sim
    \mathcal{N}\left(\mu_1+\frac{\sigma_1}{\sigma_2}\rho(x_2-\mu_2),\ (1-\rho^2)\sigma_1^2\right) \,,
\end{equation}
where $\mu_1\,(\mu_2)$ and $\sigma_1\,(\sigma_2)$ are the mean and standard deviation of $X_1\,(X_2)$, and $\rho$ is the correlation coefficient between the two.
This would be useful for deriving e.g. $\phi(p_i|p_j,z)$ from $\phi(p_i,p_j|z)$, as discussed following Eq.~\ref{eq:phi_2}.

\subsection{Fitting CGPD with GMM}

\begin{figure*}
    \centering
    \includegraphics[width=0.8\textwidth]{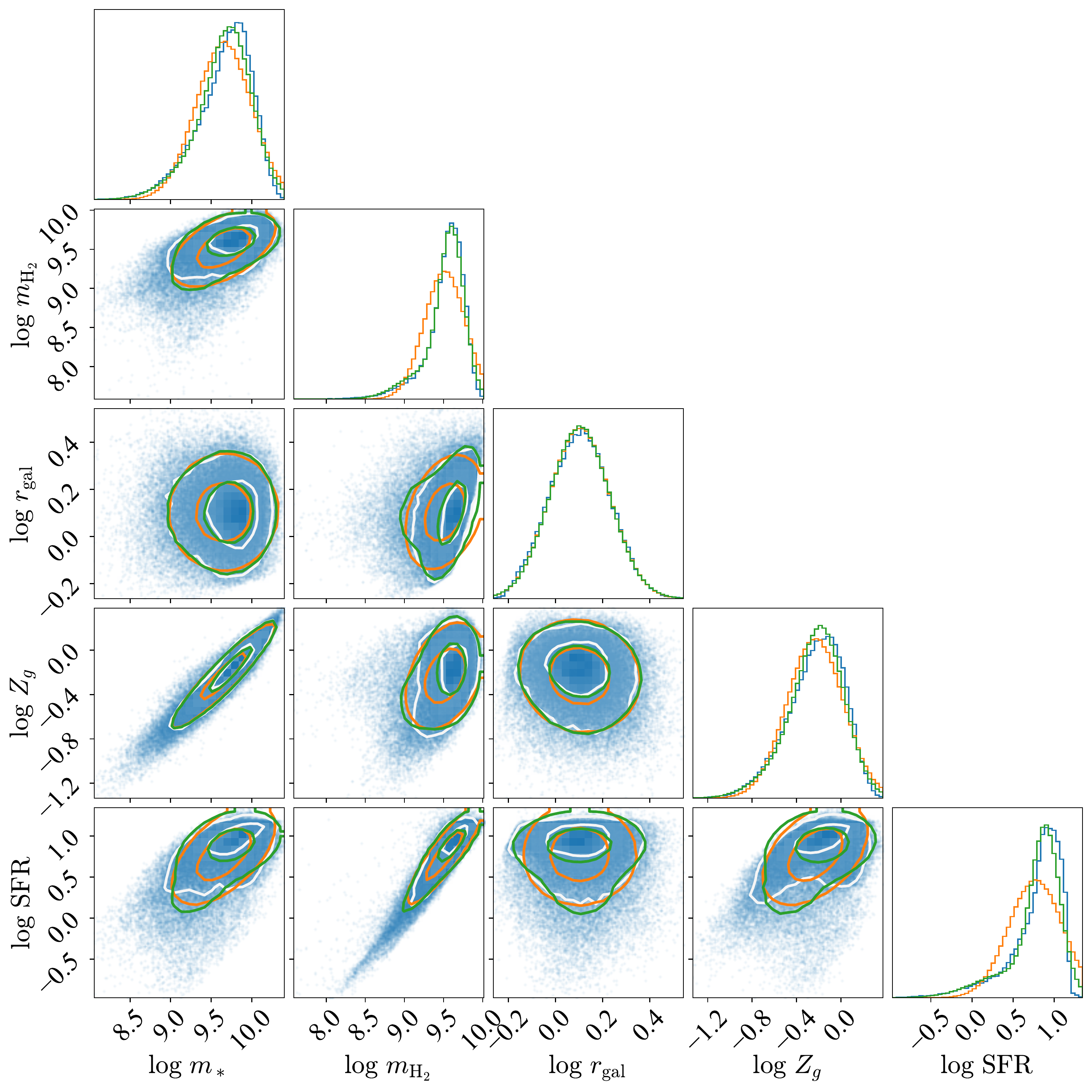}
    \caption{The blue data points with white contours denote the distribution of central galaxies hosted by halos of mass $11.4 < \log(M/\Msun) < 11.6$ at redshift $z=2$, simulated by the SAM.
    The orange and green lines denote the GMM models fitted with 1 and 2 Gaussian components respectively.
    The contours correspond to $1\,\sigma$ and $2\,\sigma$ levels in 2D, which cover $39.3\%$ and $86.5\%$ of the volume respectively.}
    \label{fig:corner_example}
\end{figure*}

Fig.~\ref{fig:corner_example} shows an example of the CGPD for central galaxies in the SAM with host halo mass $11.4 < \log(M/\Msun) < 11.6$ at redshift $z=2$, where the projected pair distribution plots are made using \texttt{corner.py}~\footnote{\url{https://github.com/dfm/corner.py}}~\citep{Foreman-Mackey2016}.
We can see that the marginal distributions for the logarithmic galaxy properties are very close to being Gaussian around the peak but with possible extended spreads on the left side.

\begin{figure*}
    \centering
    \includegraphics[width=0.8\textwidth]{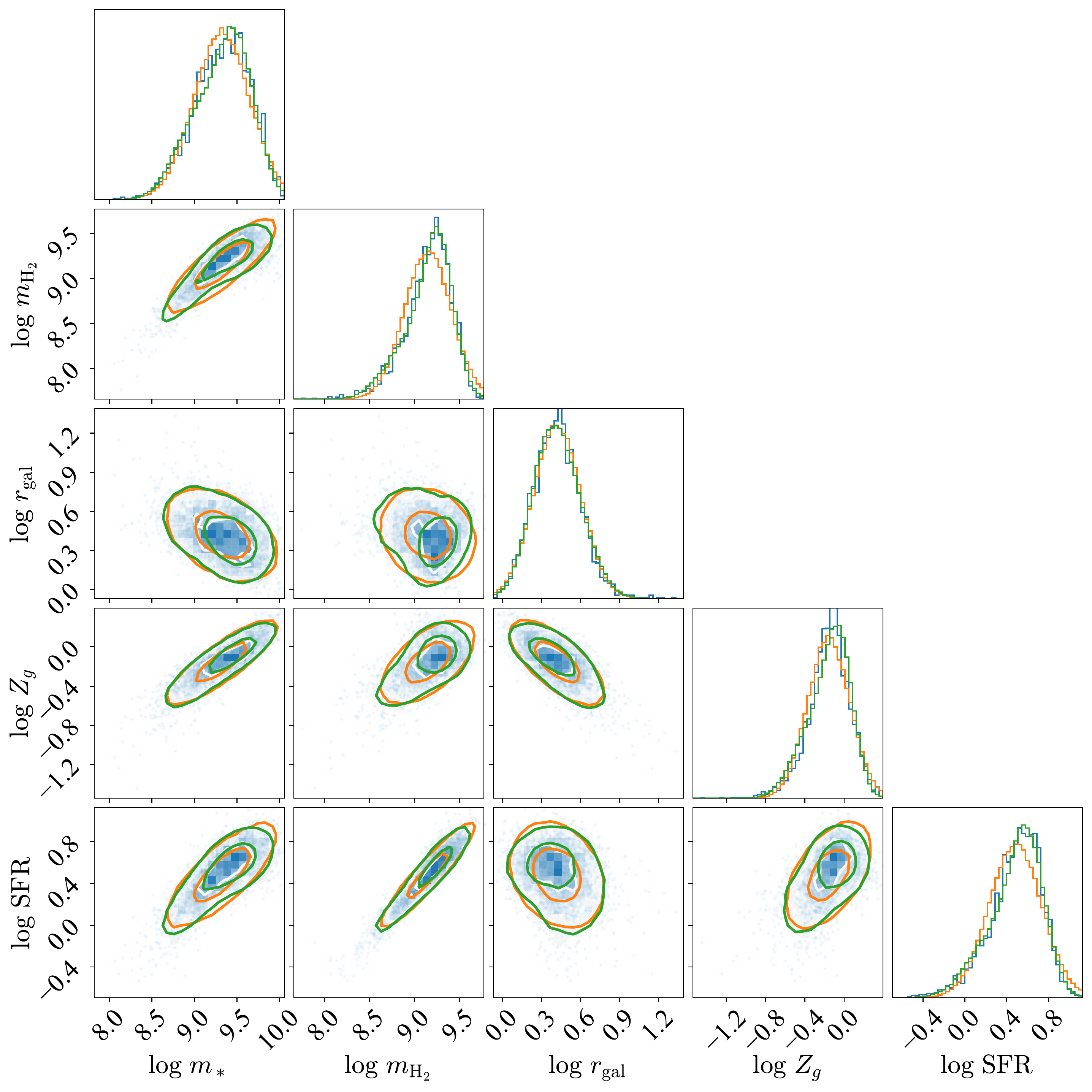}
    \caption{Similar to Fig.~\ref{fig:corner_example} but for the TNG simulation data.}
    \label{fig:corner_example_tng}
\end{figure*}
Similar to Fig.~\ref{fig:corner_example} for the SAM data, we show galaxies selected under the same criteria from the TNG simulated data in Fig.~\ref{fig:corner_example_tng}.
There are many fewer galaxies in the TNG data compared to those in the SAM. 
Although the locations and dispersions are different from those of the SAM shown in Fig.~\ref{fig:corner_example}, the distributions are still very close to being Gaussian. We conclude for now that this is likely to be a fairly robust property of CGPD in simulations and that therefore Gaussian mixtures are a promising parameterized model for this problem. 

We use GMM as a parametric model that describes the CGPD $\cgpd{\p}$, where the data vector $\bm{x}$ in Eq.~\ref{eq:gmm_def} corresponds to the galaxy property vector $\p$.
For a bin in mass and redshift $(M, z)$, given $n$ simulated galaxies $\{\p_i\}_{i=1}^n$ for fitting, the overall logarithmic likelihood function is given by
\begin{equation}
    \mathcal{L}(\{\p_i\}_{i=1}^n|\bm{\theta}) = \sum_{i=1}^n \ln\, L(\p_i|\bm{\theta}) \,,
\end{equation}
which is maximized to find the best-fit $\hat{\bm{\theta}}$ using the expectation-maximization (EM) algorithm~\citep{Dempster1977}.
In this work, we use the implementation of the GMM in \texttt{scikit-learn}~\footnote{\url{https://scikit-learn.org/stable/modules/mixture.html}}.

Given the galaxy data and GMM model, we fit the galaxies in the 5-D physical property space shown in Fig.~\ref{fig:corner_example}.
Considering the correlations between physical properties, we use the full covariance matrices for the Gaussian components.
However, this corresponds to a large number of parameters in the covariance matrix and the model might need to be further simplified in order to be used for statistical inference.
We discuss a possible dimensionality reduction based on principal component analysis (PCA) in Section~\ref{sec:dim_reduce}.

At each redshift, we divide the logarithmic host halo mass range into bins of uniform width $\Delta \log M = 0.2$.
As mentioned in Section~\ref{subsec:sam}, the halo mass resolution in our SAM simulations is $0.003\,\log\Msun$, which is much smaller than the bin size such that halo masses are well sampled in each bin.
Since different halo mass bins are fitted independently, the overall complexity of the model is also proportional to the number of bins.
However, using fewer bins means a wider spread of the distribution for each bin and possibly results in lower accuracy.
This is another trade-off depending on the usage.
Another detail to notice is that as described in Section~\ref{subsec:sam}, in our simulations from SAM based on merger trees using the Monte Carlo method, the same number of halos are generated for the one thousand discrete halo masses.
Therefore for wide halo mass bins, to be realistic it would be necessary to subsample the halos within a bin to match the HMF.

The model complexity of a GMM is proportional to the number of Gaussian components $N_{\rm gc}$, which is an important hyper-parameter that needs to be determined.
A typical model selection is based on the Bayesian information criterion (BIC), which attempts to achieve a balance between the complexity and performance of the model,
\begin{equation}
    {\rm BIC} = {\rm dim}(\bm{\theta})\ln n - 2\hat{\mathcal{L}} \,,
\end{equation}
where ${\rm dim}(\bm{\theta})$ denotes the total number of parameters in the model, $n$ is the number of data points, and $\hat{\mathcal{L}}$ is the log-likelihood at the best-fit $\hat{\bm{\theta}}$.
For the same data, the models with lower BIC scores are preferred.
\begin{figure}
    \centering
    \includegraphics[width=\columnwidth]{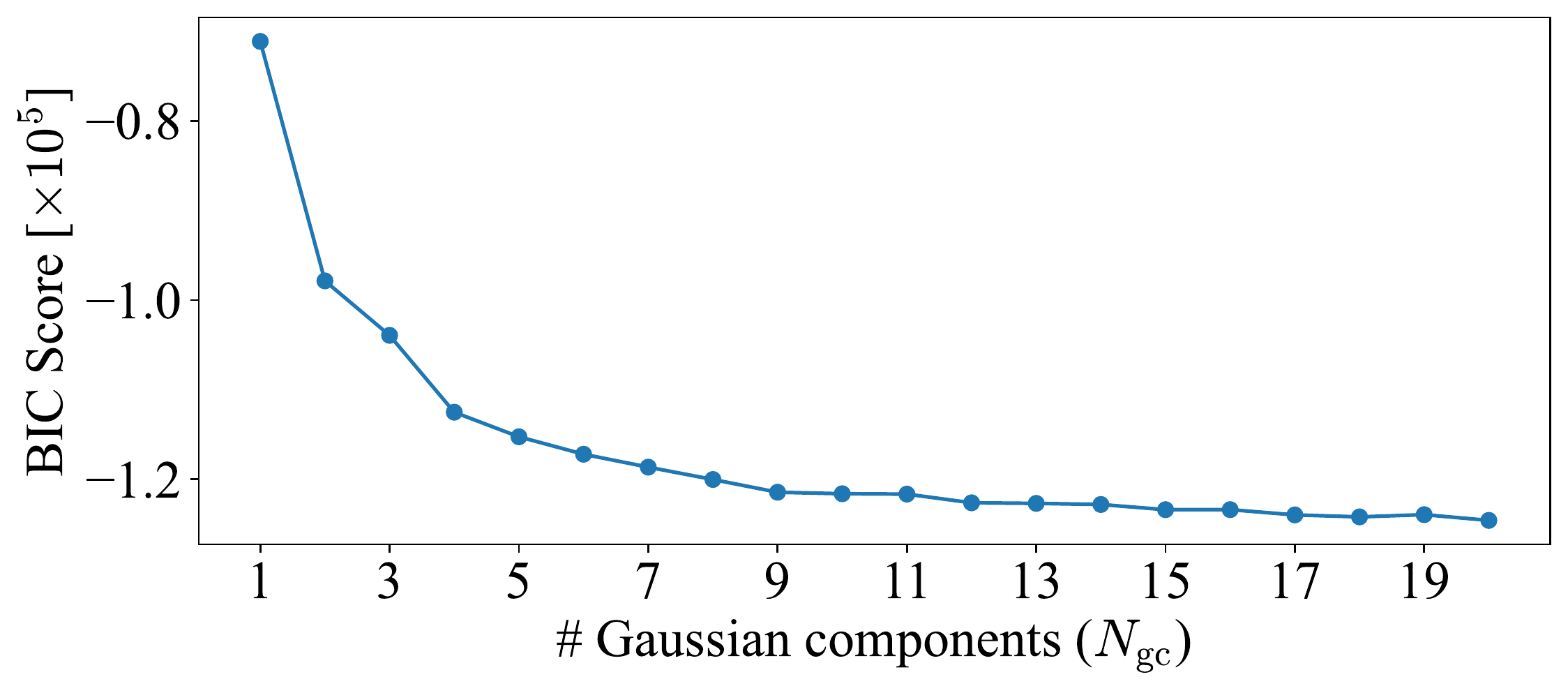}
    \caption{BIC scores for different numbers of components in a GMM, using the same galaxy sample from the SAM as that shown in Fig.~\ref{fig:corner_example}.}
    \label{fig:bic_example}
\end{figure}
In Fig.~\ref{fig:bic_example}, for GMM fitted on the same example group of galaxies shown in Fig.~\ref{fig:corner_example}, we show the BIC scores for models with different numbers of components ($N_{\rm gc}$).
The BIC score decreases as we increase $N_{\rm gc}$ starting from just one component, and the curve flattens very rapidly.
Similar curves are also observed for galaxies at other redshifts and other halo mass bins.
These curves indicate that a small number of components is sufficient, and continuing to increase $N_{\rm gc}$ to a very large value (e.g. $>\,10$) will not significantly improve the fitting performance, but will only make the model more complicated.
The BIC score is a general criterion for model selection based on the likelihood, while in practice the trade-off between accuracy and model complexity depends on the specific usage.
If we want to use the model as an emulator of the SAM simulations, then we may want to use more components to make it more accurate.
On the other hand, if we want to use the model for statistical inference, we would prefer to make the model as simple as possible while achieving acceptable accuracy.
In our main analyses, we always start with very few (i.e. 1, 2 or 3) components.

\subsection{Model Performance}

To check how well the GMM fits the SAM and TNG data, for some of the galaxy property functions mentioned in Section~\ref{sec:cgpd}, we compare the results given by the best-fit GMM with those computed directly from the data.

\begin{figure*}
    \centering
    \includegraphics[width=\columnwidth]{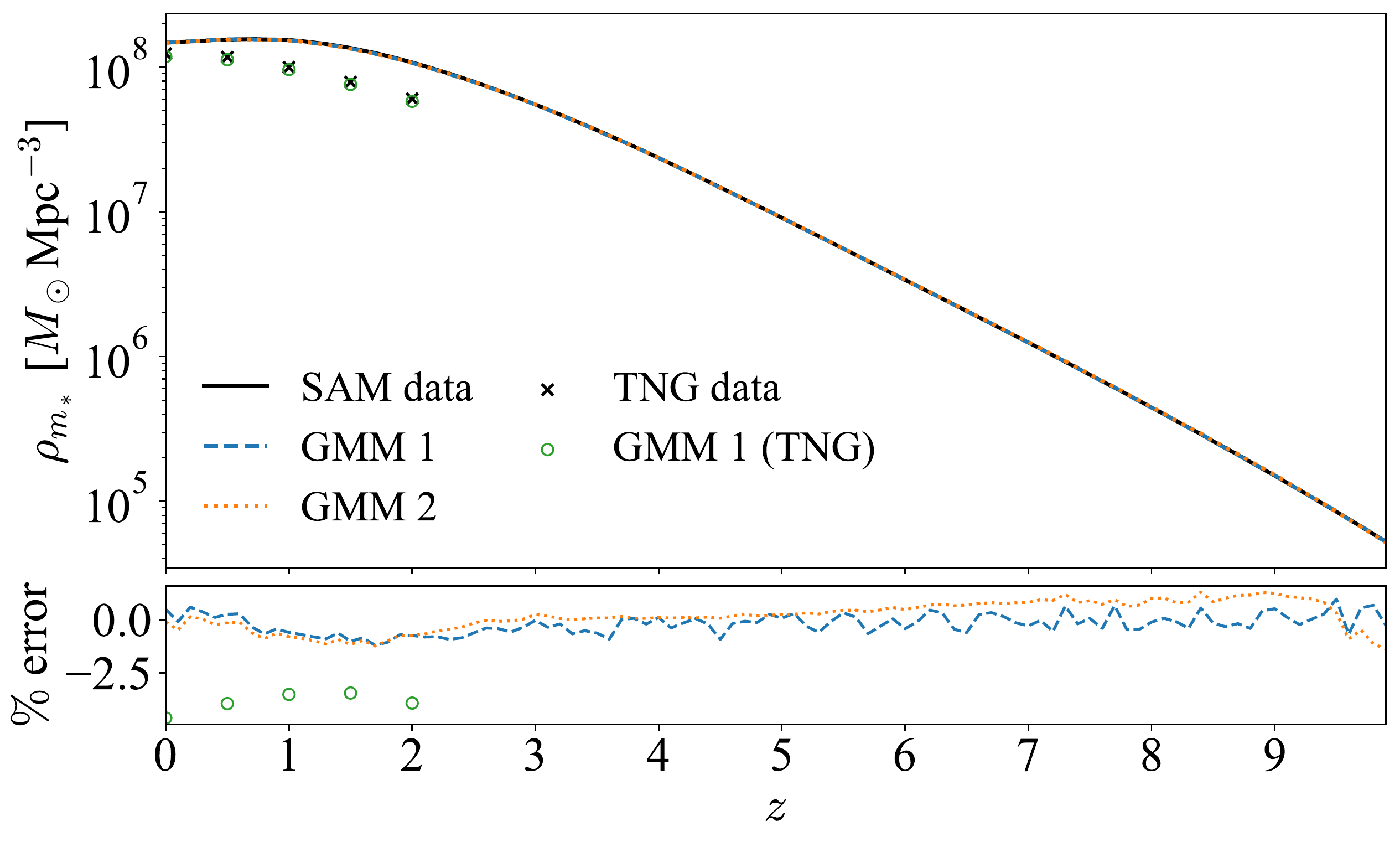}
    \includegraphics[width=\columnwidth]{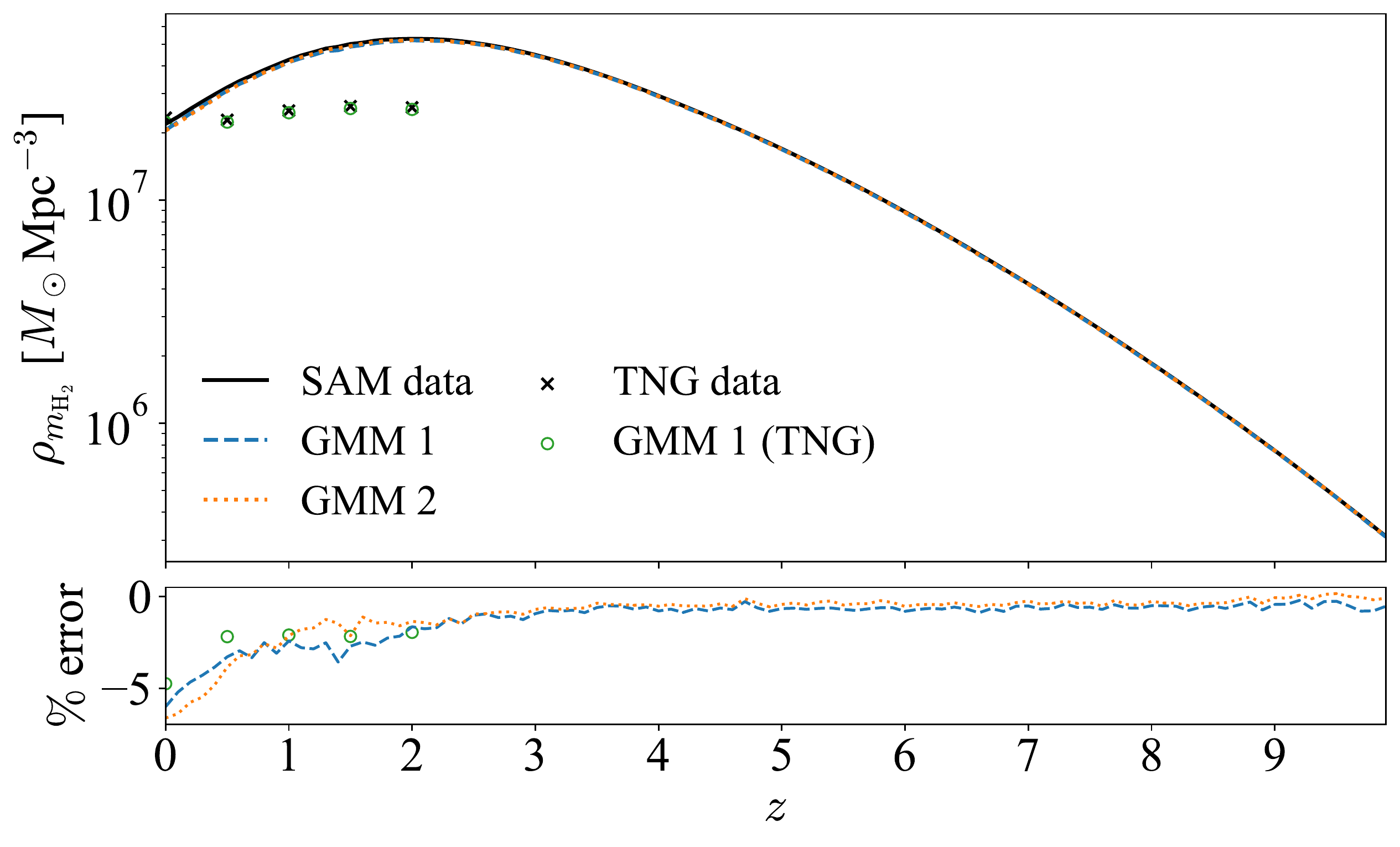} \\
    \includegraphics[width=\columnwidth]{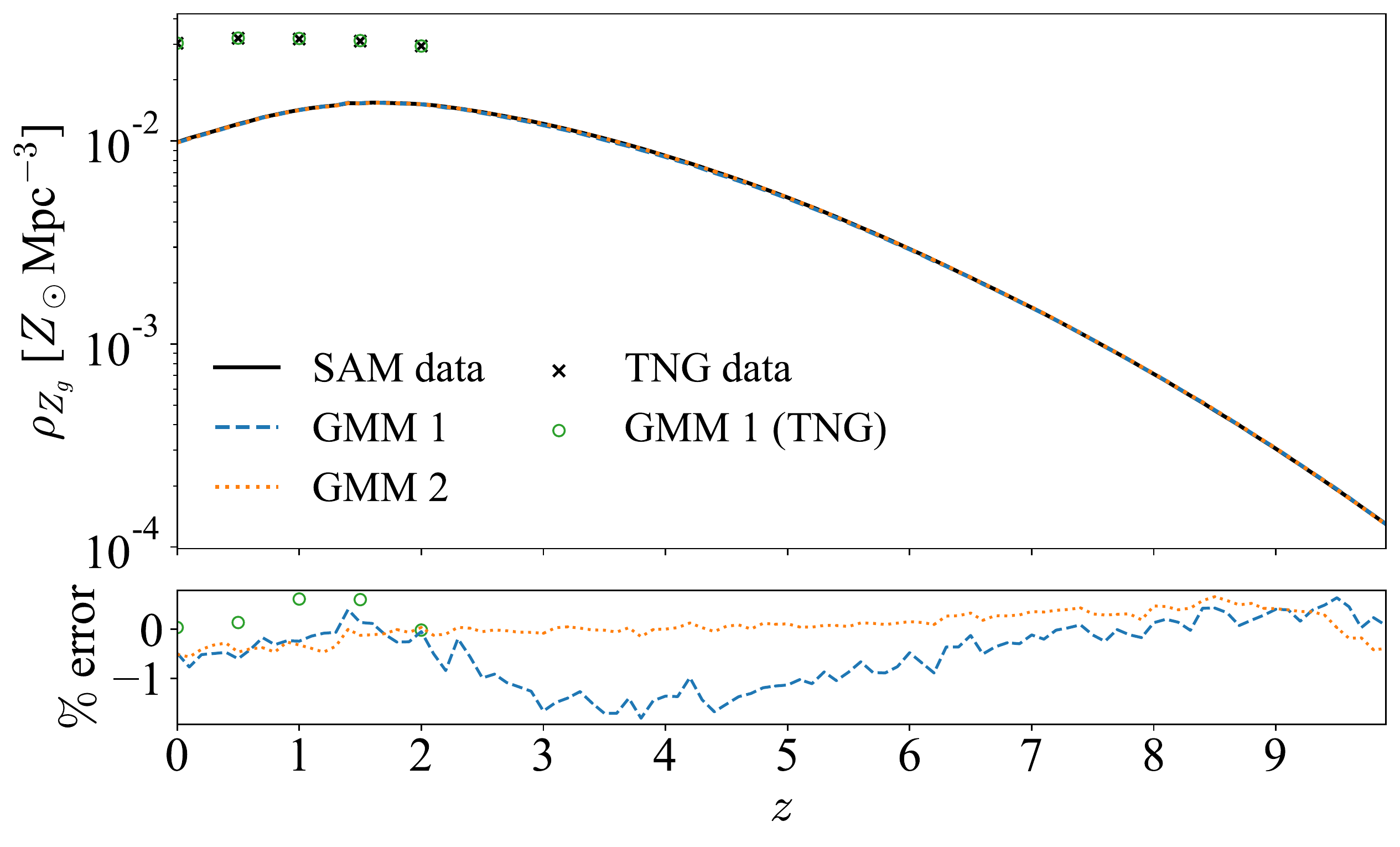}
    \includegraphics[width=\columnwidth]{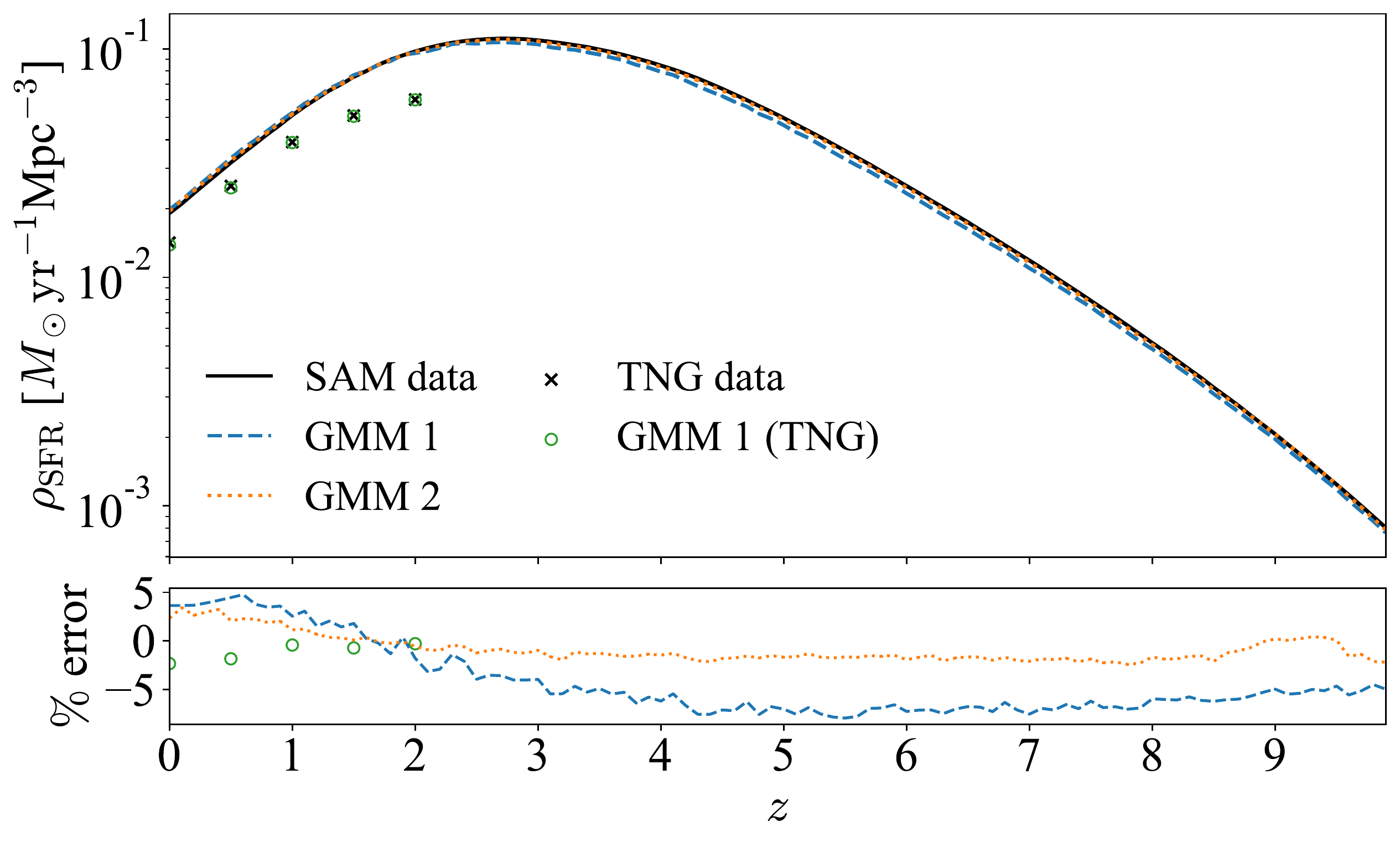} \\
    \caption{Cosmic densities of the galaxy physical properties.
    The solid blue lines are computed directly from the SAM simulation data.
    The dashed blue (dotted orange) lines denote the best-fit GMM with one (two) component(s).
    The lower panels show the percent errors of the GMMs compared to the SAM data.
    We also show the TNG simulated data at five redshifts (see Section~\ref{subsec:tng}), along with the best-fit GMM with one component and the corresponding percent error.}
    \label{fig:prop_cosmic_den}
\end{figure*}

First we consider the cosmic densities of galaxy properties defined in Eq.~\ref{eq:prop_cosmic_den}.
In Fig.~\ref{fig:prop_cosmic_den}, we show the results for GMM with one or two components for the one hundred uniformly spaced redshift bins in $0\leq z < 10$.
Also shown are the TNG data at five available redshifts (see Section~\ref{subsec:tng}) and the corresponding best-fit GMMs with just one component.
For both SAM and TNG simulated data, with just one or two Gaussian components, the percent errors of the cosmic densities evaluated with the GMM compared to the results computed directly from the simulation are mostly within $5\,\%$, which is acceptable given the uncertainties in state-of-the-art observational constraints.
For these integrated quantities, the improvement obtained when using one more component which can capture the tails in distributions (see 1-D histograms in Fig.~\ref{fig:corner_example}) is not very significant.

We also consider the galaxy one-property distribution functions defined in Eq.~\ref{eq:phi_1}.
\begin{figure*}
    \centering
    \includegraphics[width=\columnwidth]{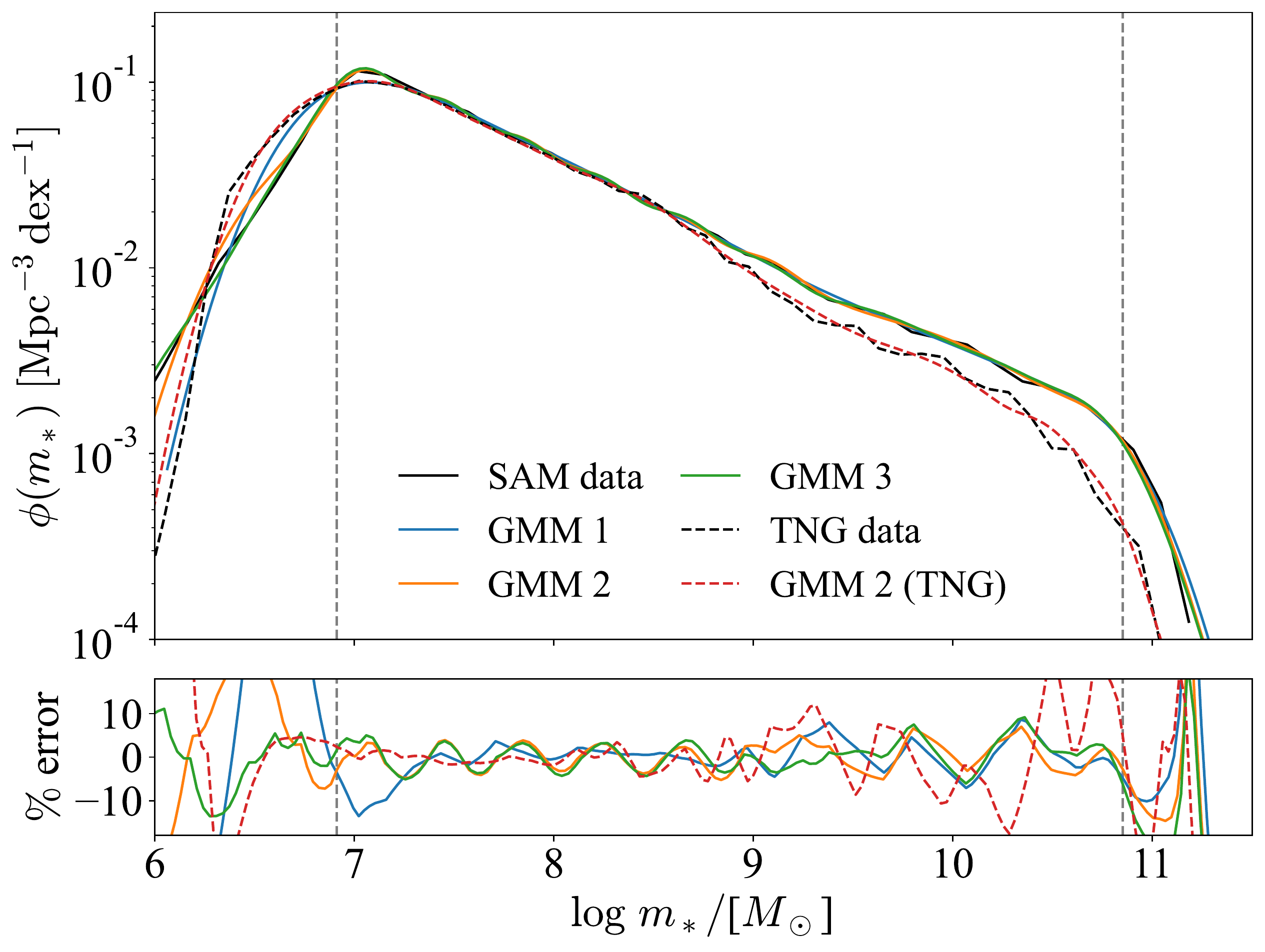}
    \includegraphics[width=\columnwidth]{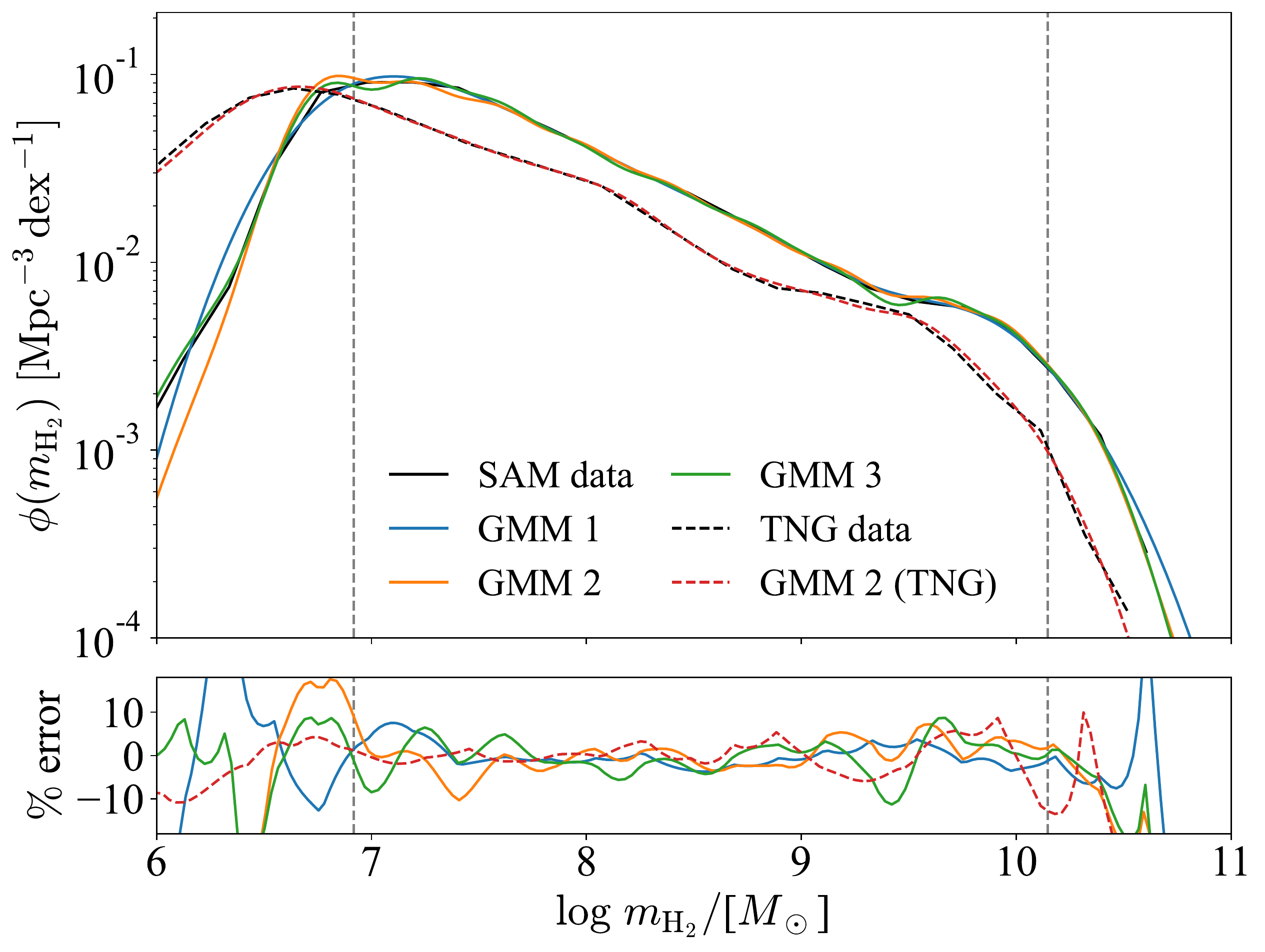} \\
    \includegraphics[width=\columnwidth]{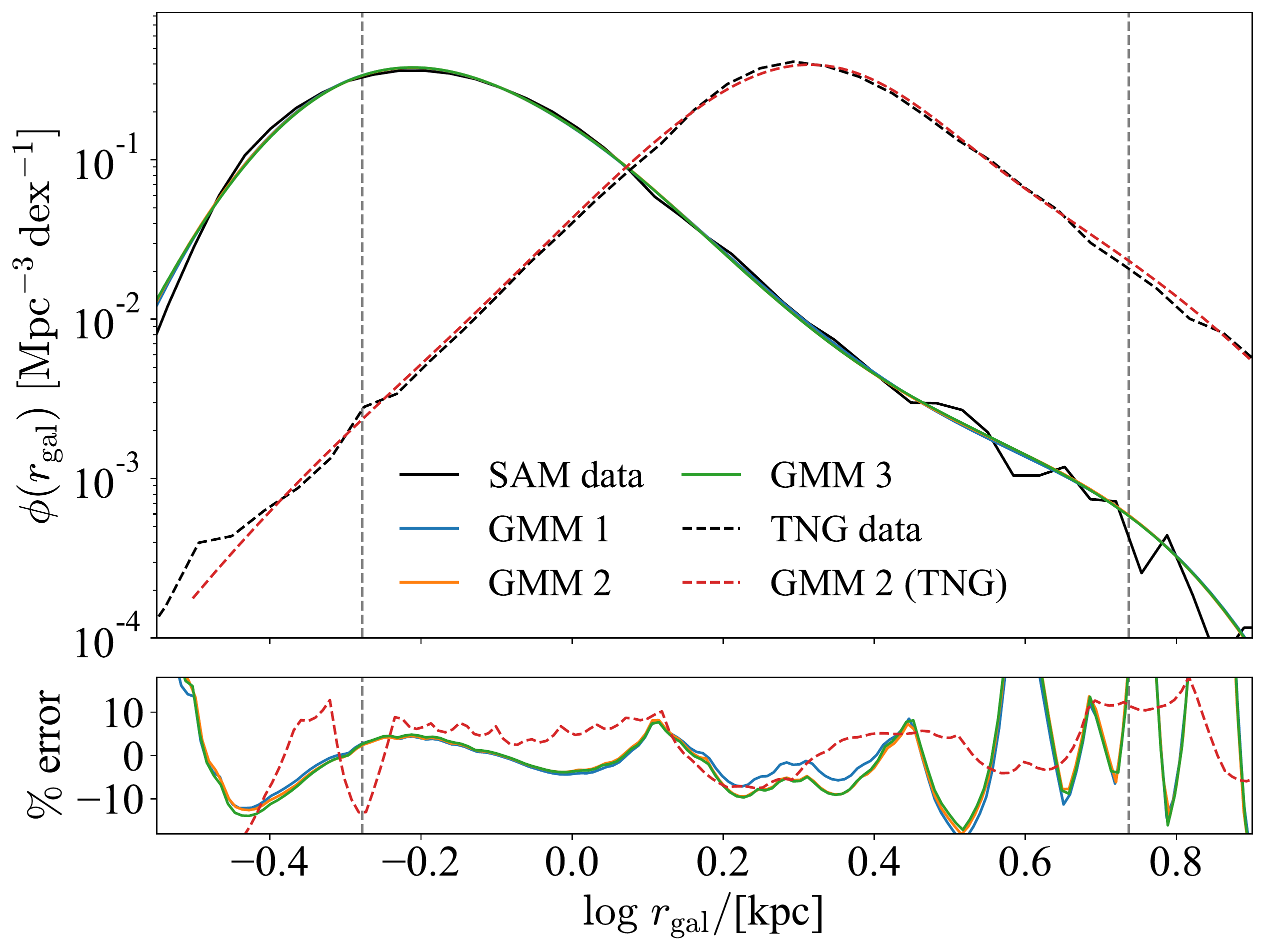}
    \includegraphics[width=\columnwidth]{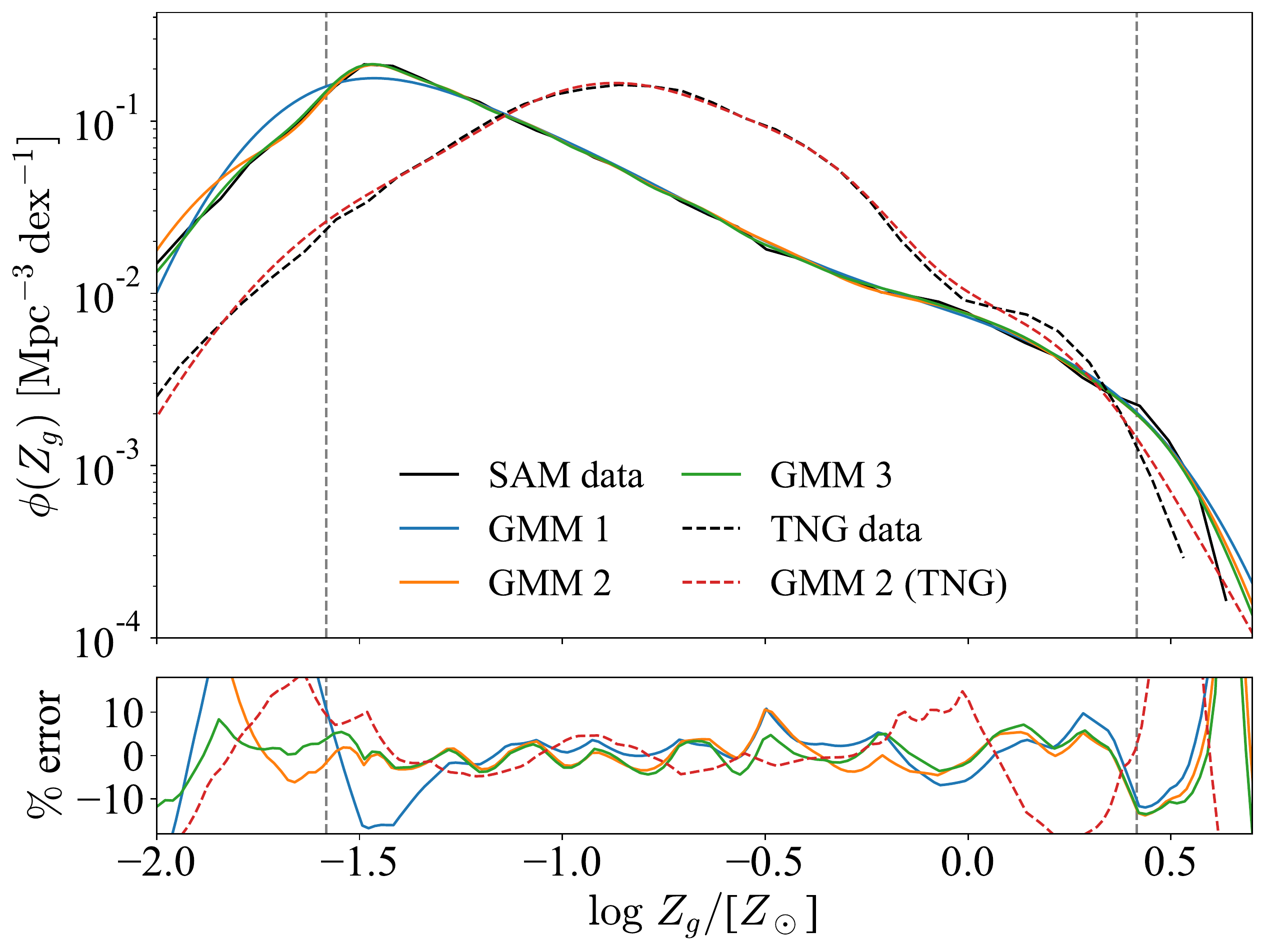} \\
    \includegraphics[width=\columnwidth]{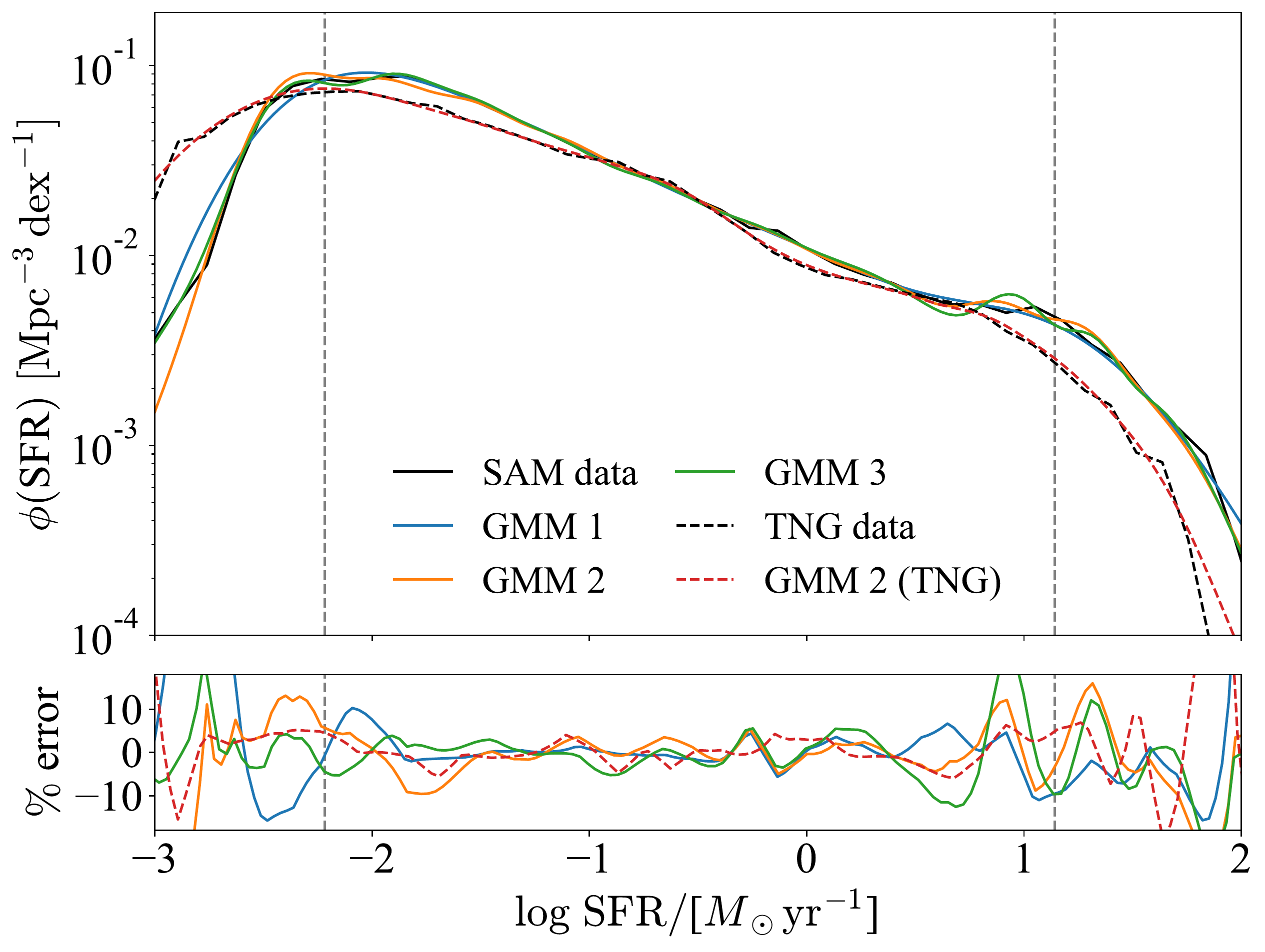}
    \caption{The galaxy one-property distribution functions defined in Eq.~\ref{eq:phi_1}.
    We show the stellar mass function (\textit{upper left}), molecular gas mass function (\textit{upper right}), galaxy radius function (\textit{middle left}), gas phase metallicity function (\textit{middle right}), and SFR function (\textit{lower}) at redshift $z=2$.
    The black line is the function computed directly from the SAM data.
    The blue, orange and green lines show the function obtained from the best-fit GMM using 1, 2 and 3 components respectively for each halo mass bin. We show the TNG results and its best-fit GMM with 2 components with dashed lines.
    The corresponding percent errors compared to the SAM and TNG simulation data are shown in the lower panels.
    Since we consider halo masses in $10 \leq \log(M/\Msun) \leq 13$, there are also limits outside which the galaxies hosted by less or more massive halos may not be negligible.
    We denote these limits with grey dashed vertical lines, which are the mean properties of the galaxies in the lowest and highest mass bins we consider, i.e. $10 \leq \log(M/\Msun) \leq 10.2$ and $12.8 \leq \log(M/\Msun) \leq 13$.
    }
    \label{fig:gal_func_one}
\end{figure*}
In Fig.~\ref{fig:gal_func_one}, for both the SAM and TNG, we show these functions at redshift $z=2$ as an example, including the function computed directly from the simulation and the best-fit GMMs with different numbers of components.
GMM with just one component can already capture most of the information, especially the main slope of the function.
Compared with the cosmic densities above, using more than one component could help capturing more information.
In general, using no more than 3 components can already reproduce the galaxy one-property functions to very high accuracy.

We now consider the galaxy two-property distribution functions (scaling relations) defined in Eq.~\ref{eq:phi_2}.
\begin{figure*}
    \centering
    \includegraphics[width=\columnwidth]{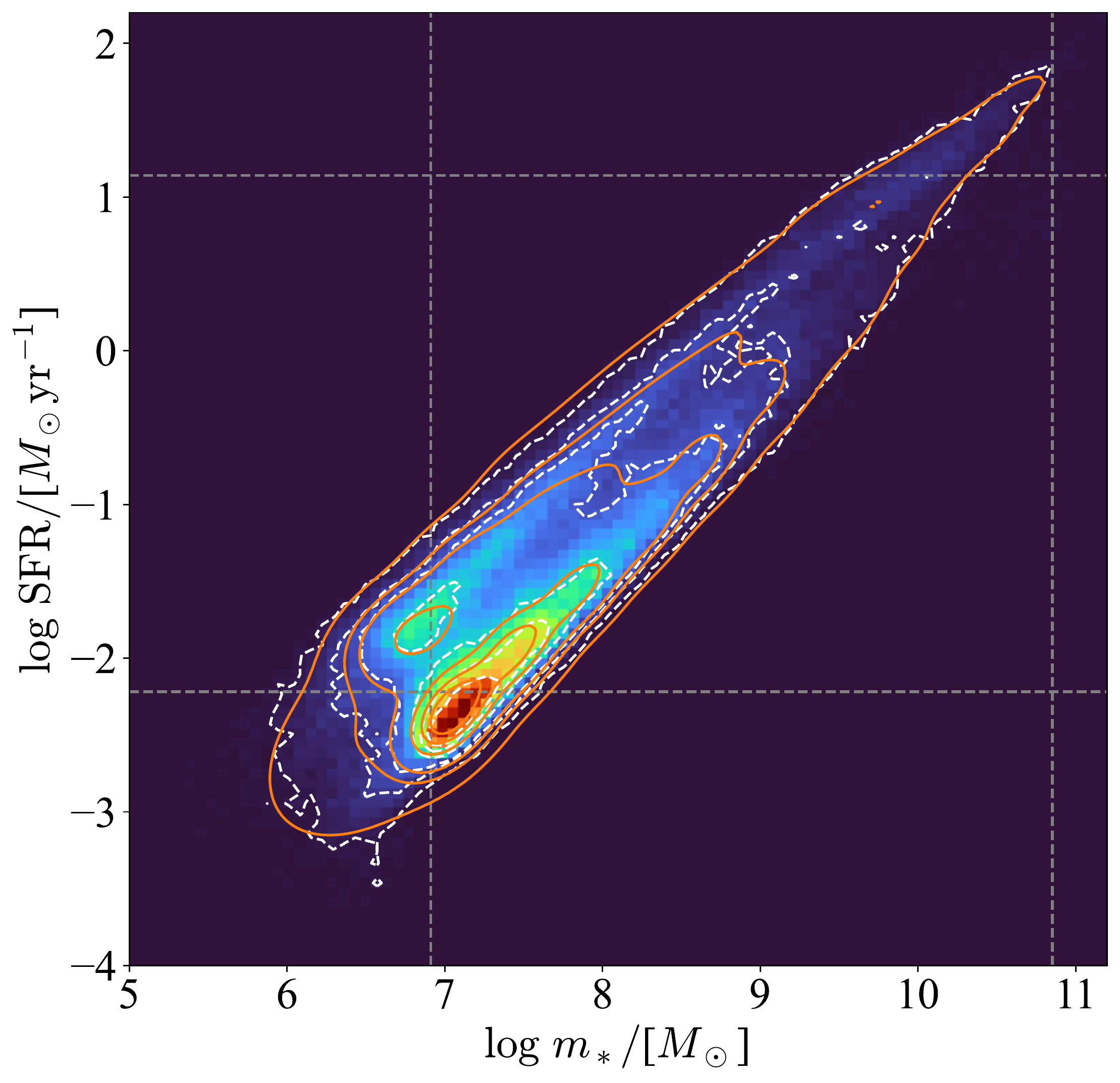}
    \includegraphics[width=\columnwidth]{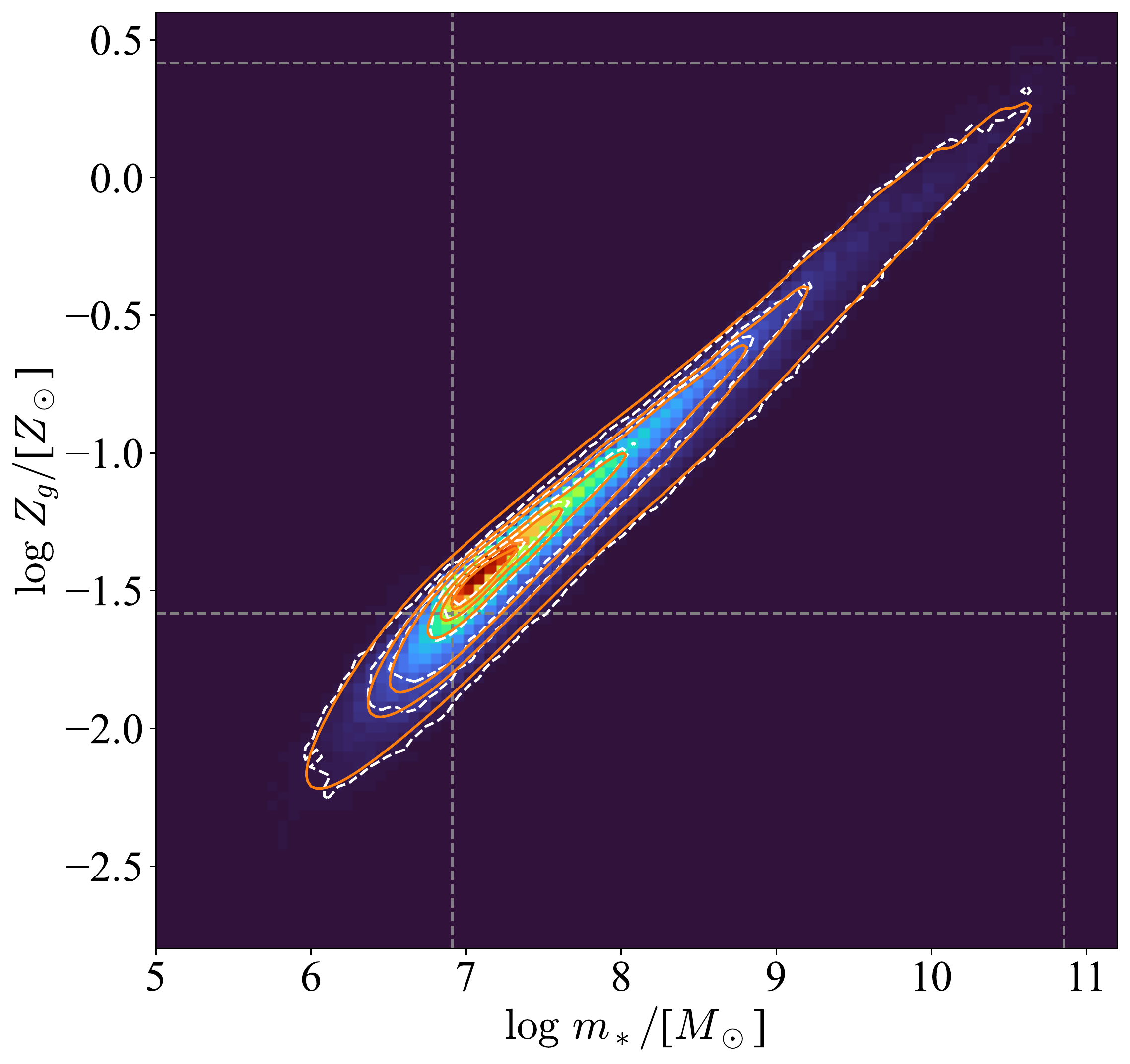} \\
    \includegraphics[width=\columnwidth]{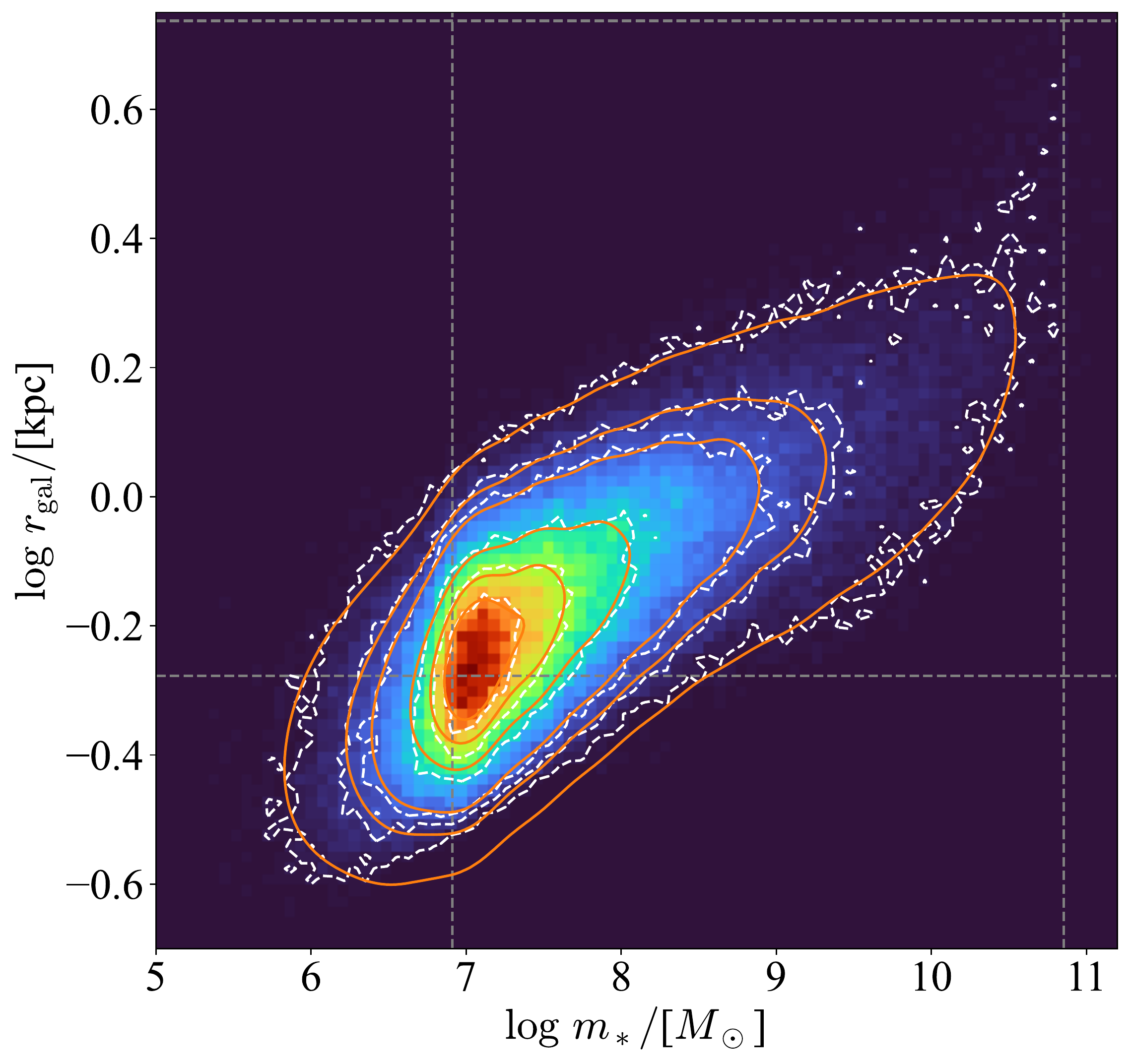}
    \includegraphics[width=\columnwidth]{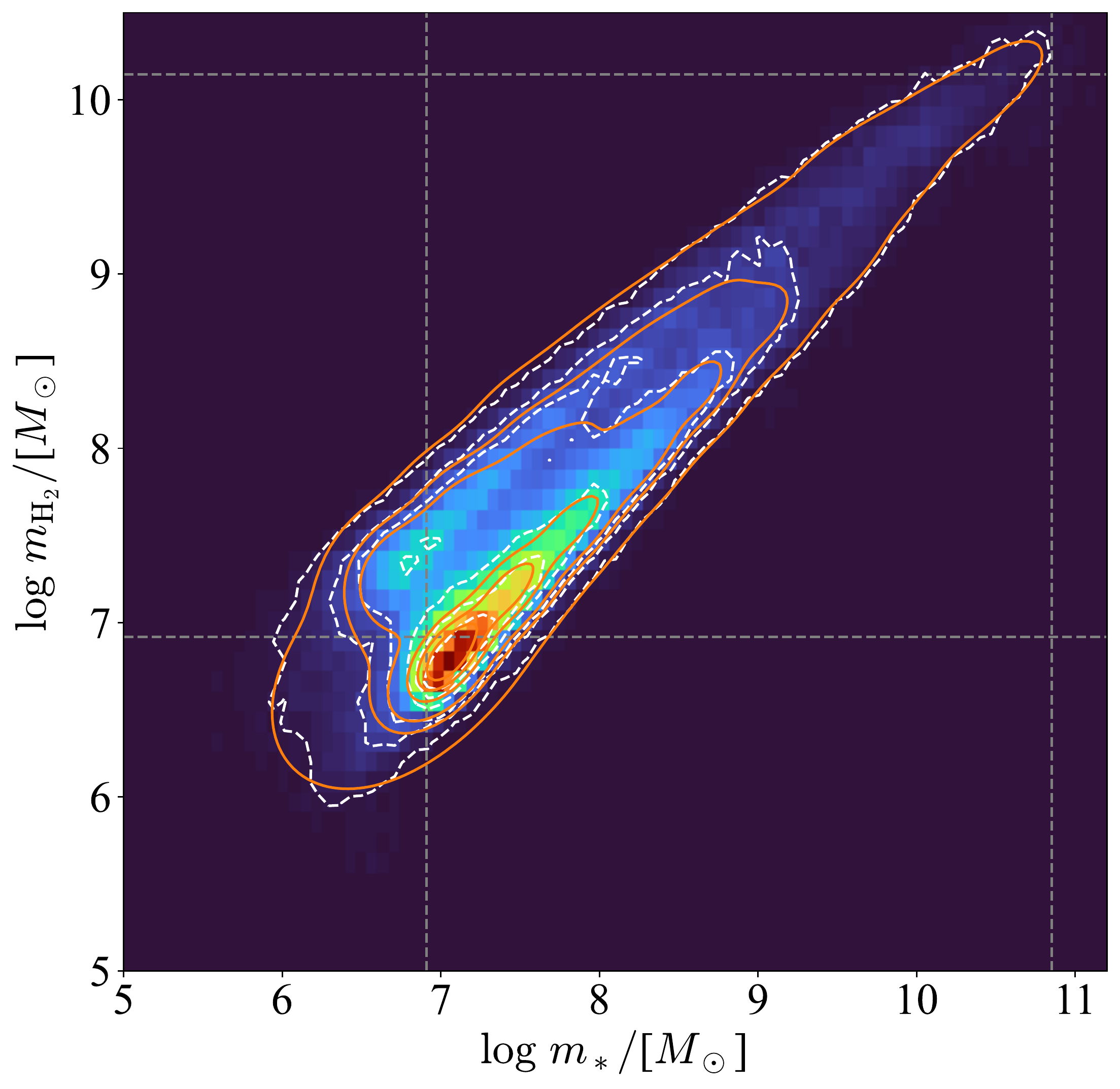}
    \caption{Galaxy two-property distribution functions defined in Eq.~\ref{eq:phi_2}.
    Here we show $\phi(m_*,\sfr)$ (\textit{upper left}), $\phi(m_*,Z_g)$ (\textit{upper right}), $\phi(m_*,\rgal)$ (\textit{lower left}) and $\phi(m_*,\mhh)$ (\textit{lower right}) at redshift $z=2$.
    The histograms with white-dashed contours show the SAM simulation results.
    The orange contours show the best-fit GMM with three components.
    The contours are plotted at (0.7, 0.5, 0.3, 0.1, 0.05, 0.01) times the maximum value of the corresponding distribution.
    The vertical and horizontal grey dashed lines denote the same limits of each property as in Fig.~\ref{fig:gal_func_one}.}
    \label{fig:gal_func_two}
\end{figure*}
\begin{figure*}
    \centering
    \includegraphics[width=\columnwidth]{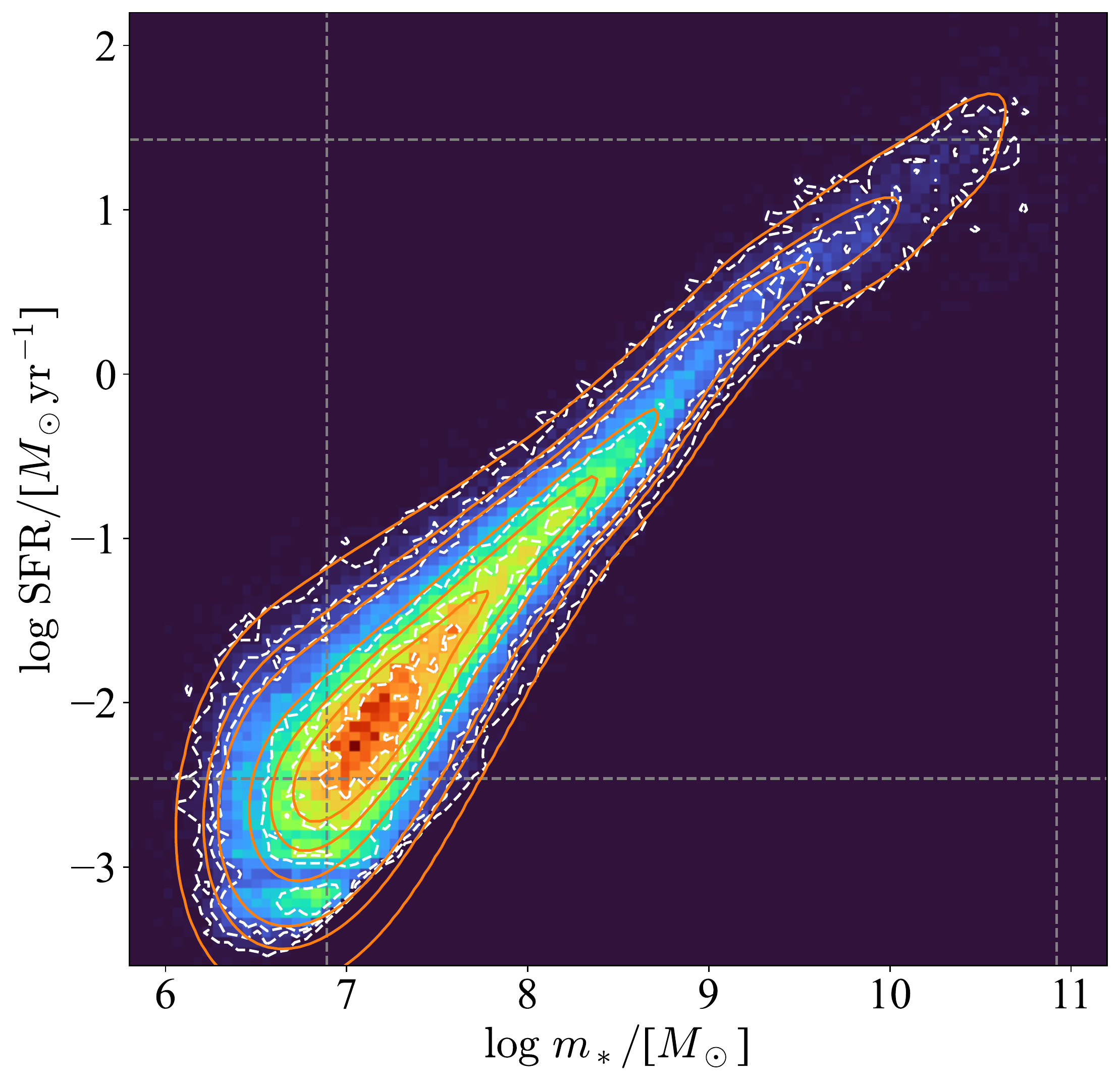}
    \includegraphics[width=\columnwidth]{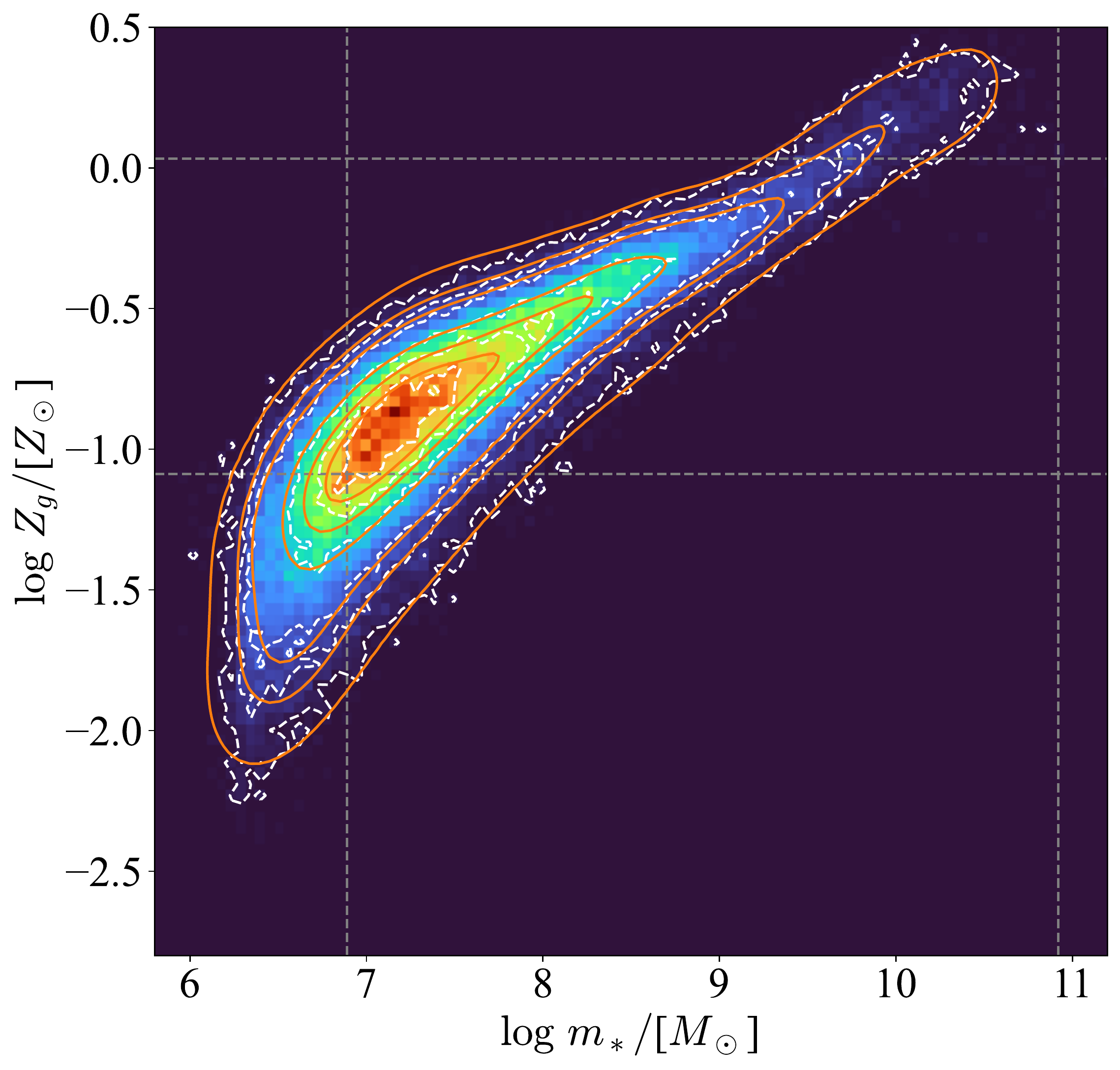} \\
    \includegraphics[width=\columnwidth]{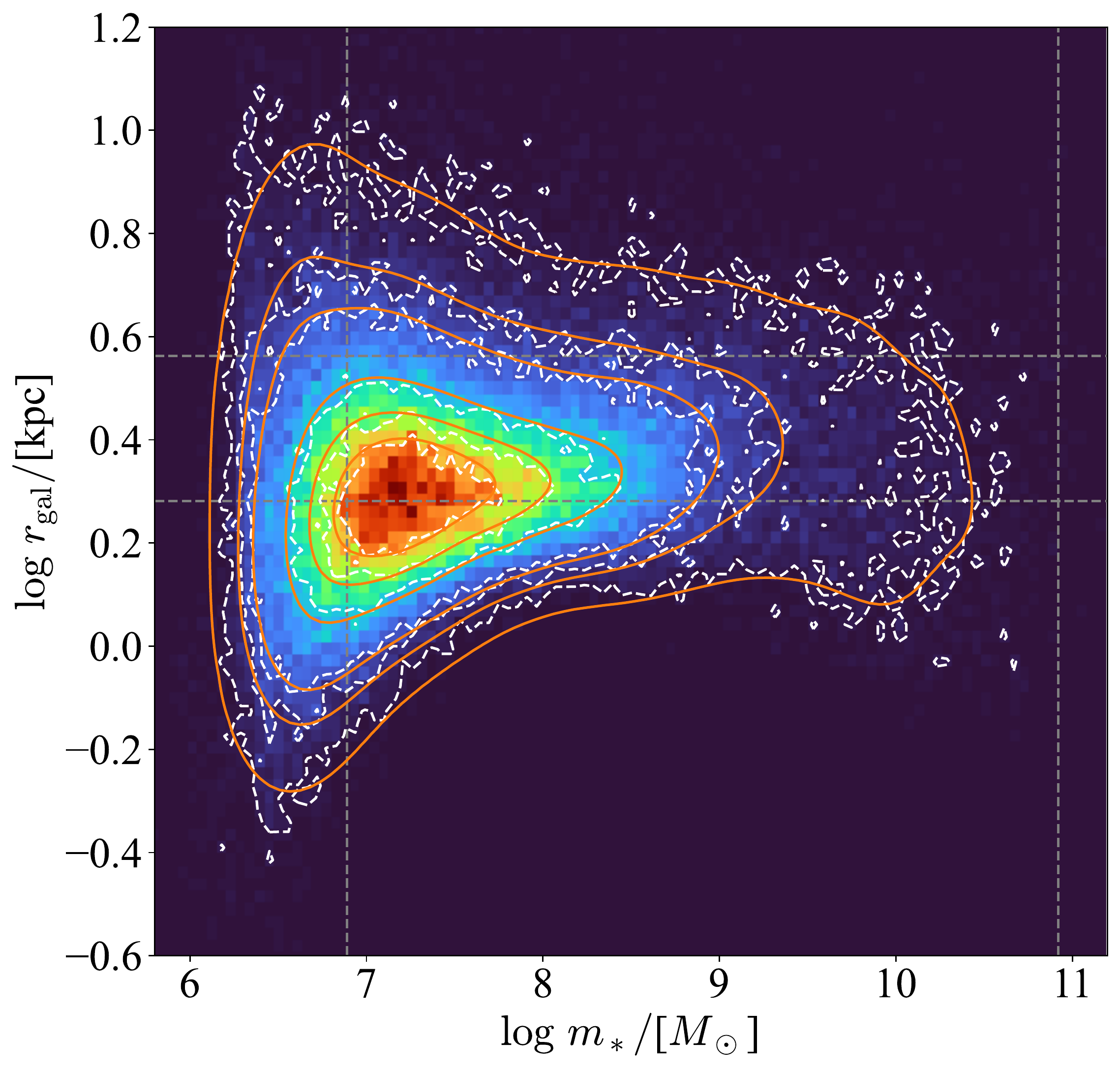}
    \includegraphics[width=\columnwidth]{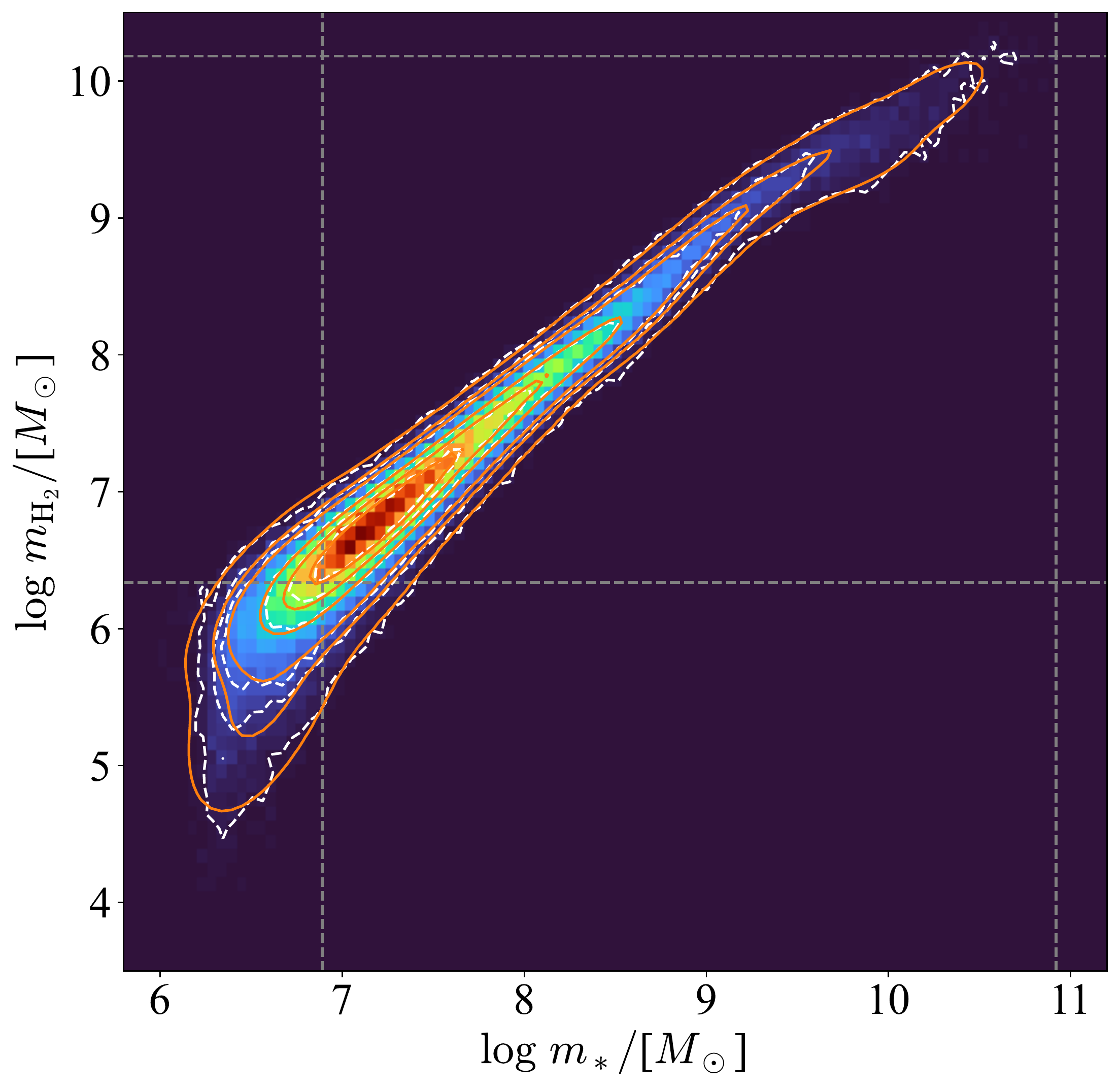}
    \caption{Similar to Fig.~\ref{fig:gal_func_two} but for the TNG simulation data.}
    \label{fig:tng_gal_func_two}
\end{figure*}
For example, we plot $\phi(m_*,\sfr)$, $\phi(m_*,Z_g)$, $\phi(m_*,\rgal)$, and $\phi(m_*,\mhh)$ at redshift $z=2$ in Fig.~\ref{fig:gal_func_two} for SAM and Fig.~\ref{fig:tng_gal_func_two} for TNG.
For clarity, here we only show a GMM with three components, which can already reproduce the data distribution very well.
Using fewer (i.e. one or two) components could also be acceptable depending on how many details we would like to recover.
Scaling relations can also be observed in some of the distributions.

\subsection{Dimensionality Reduction: Principal Component Analysis} \label{sec:dim_reduce}

As expected and also shown in Fig.~\ref{fig:corner_example}, there are strong correlations between the galaxy physical properties we consider.
This makes it possible to reduce the dimension of the data without losing much information using linear transformations like Principal Component Analysis.
With the data standardized (i.e. shifted to zero mean and normalized to unit variance) on each dimension, starting from the principal component (PC) with the largest variance, PCA sequentially finds the next PC with the next largest variance while being linearly independent of the previous PCs.
Lower data dimensionality could reduce the number of parameters needed in the model and make it more feasible for use in an inference pipeline.
Besides, the linear independence (i.e. zero covariances) between PCs makes it possible to simplify certain models like a multivariate Gaussian distribution even further.

\begin{figure*}
    \centering
    \includegraphics[width=\textwidth]{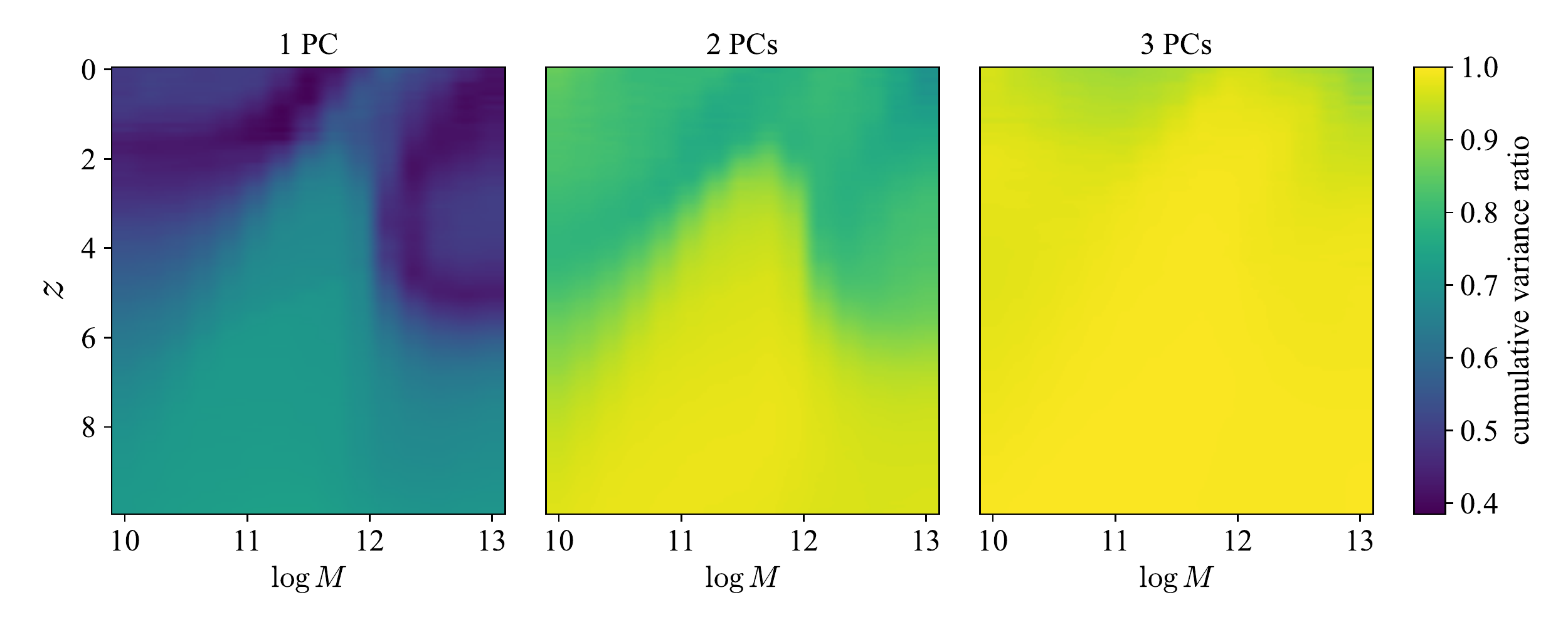}
    \caption{For the distribution of galaxies in each $(M, z)$ bin, cumulative percentages of variance explained by PCs with up to the first, second and third largest variance.
    }
    \label{fig:var_ratio_pc}
\end{figure*}

\begin{figure}
    \centering
    \includegraphics[width=\columnwidth]{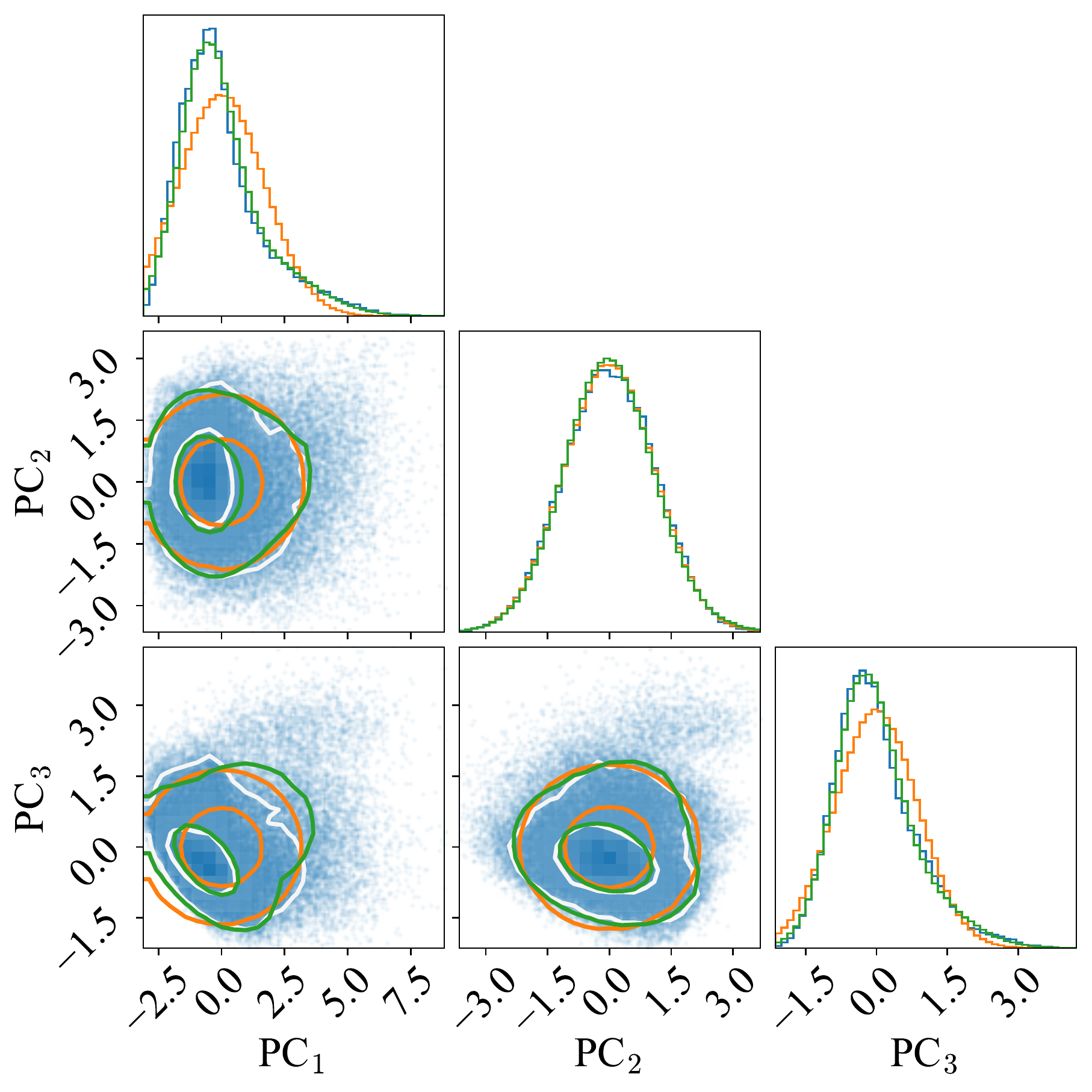}
    \caption{Blue data points with white contours show the distribution of the same group of galaxies shown in Fig.~\ref{fig:corner_example} in the space of the first 3 PCs.
    The orange and green lines denote the distributions with GMM using 1 and 2 Gaussian components respectively.
    The contours correspond to $1\,\sigma$ and $2\,\sigma$ levels in 2D, which cover $39.3\%$ and $86.5\%$ of the volume respectively.}
    \label{fig:corner_example_pc}
\end{figure}

Here we perform PCA on central galaxies for each $(M,z)$ bin.
The cumulative percentages of variance explained by PCs up to the third largest variance is shown in Fig.~\ref{fig:var_ratio_pc}.
It turns out that using just 3 PCs can already recover more than 90 percent of the variance.
The distribution in the space of first 3 PCs for the galaxies in Fig.~\ref{fig:corner_example}, is shown in Fig.~\ref{fig:corner_example_pc}.
Since PCA attempts to minimize parameter covariance, we do not see very strong correlations as we have seen in Fig.~\ref{fig:corner_example} for the physical properties.
However, it is still possible for PCs to have nonlinear correlations.

Therefore with acceptable information loss, instead of in the 5-D physical property space, we can describe the distribution in the 3-D PC space and convert it back to the physical property space for usage.
An additional advantage of using GMM is that a Gaussian distribution can also be analytically transformed between two spaces.
Each Gaussian component fitted in the PC space also corresponds to a Gaussian component in the physical property space.
With the mapping between the PC vector $\bm{y}$ and physical property vector $\p$ given by the linear transformation $\bm{y} = \bm{W}\bm{p}$ with $\bm{W}$ being a 2-D matrix of dimension ${\rm dim}(\bm{y}) \times {\rm dim}(\bm{p})$, the covariance matrices are related by
\begin{equation}
    \cov(\bm{p}) = \bm{W}^T\cov(\bm{y})\bm{W} \,.
\end{equation}
In Fig.~\ref{fig:corner_example_pc}, we show the best-fit GMM with one and two components in the PC space.
For just one component, the covariance matrix is diagonal since there is no covariance.
For two components, since each of them only cover part of the data, the full covariance matrices need to be considered.
If we convert the best-fit GMM back to the physical property space, the models are show a similar level of agreement as Fig.~\ref{fig:corner_example} where we fit the physical properties directly.

Fitting the model in PC space does provide the potential to simplify the model.
However, the transformation matrix between PC and physical property spaces also introduces additional parameters.
We leave further exploration for a future work.

\section{Discussion} \label{sec:discussion}
In this section, we discuss the significance of the work presented here, as well as some of the caveats and limitations of this analysis. 

\subsection{Significance of the CGPD for understanding galaxy formation}
The CGPD is a natural extension of traditional halo models, in that it links halo mass and redshift to galaxy properties. The new feature is that, while traditional halo models attempt to quantify this mapping with a single galaxy property at a time, and typically do not model the covariance between galaxy properties, the CGPD describes the mapping to an n-dimensional vector of galaxy properties and explicitly includes the covariances. We note that what appears as scatter or dispersion in a traditional style halo mass vs. single galaxy property plot is in some cases actually a result of projecting a higher dimensional manifold into a plane. Thus we expect the CGPD approach to partially, though perhaps not fully, also provide a natural description of dispersion along with the correct residual correlations. The CGPD as extracted from physics-based simulations can be an interesting vehicle for comparing different simulations or studying the impact of varying physical processes. This has not been a focus of this work, but it is notable that while the CGPD for the Santa Cruz SAM and TNG appear qualitatively quite similar, some of the quantities differ in the location of the mean and also in the width of the distributions. It would be interesting to systematically study the impact of varying physical processes and/or cosmology on the CGPD. Better understanding why there is a relatively ``thin" manifold in the n-dimensional property space and how physical processes shape the location and thickness of this manifold is also an interesting open question. 

\subsection{Limitations of current modeling approach}
\subsubsection{Star forming and quenched galaxies}

For practical reasons, we have only included central, star-forming galaxies in this analysis. This choice should not have a large impact on the accuracy of our method for our main target application, which is line intensity mapping. The lines of interest are produced by star-forming galaxies. For other applications, including the full galaxy population (not only star-forming galaxies) would certainly lead to a more multi-modal CGPD, but we are confident that this could still be modeled with a GMM approach using more components \citep[see e.g.][]{2019ApJ...872..160H}.

\subsubsection{CGPD for satellite galaxies} \label{sec:satellites}
According to the SAM predictions, the overall cosmic volume densities and other integrated quantities of the galaxy properties are dominated by central galaxies, with the contributions from satellites mostly less than $10\,\%$. However, it would also be straightforward to include a separate CGPD for satellite galaxies. 
For satellite galaxies, instead of the host mass $M$, it is customary to use a quantity like $M_{\rm max, sub}$, which is defined as the maximum mass that the host halo of the satellite galaxy had before it became a satellite \citep[e.g.][]{Behroozi2019}. Under the (common) ansatz that the properties of satellite galaxies depend only on the \emph{sub-halo} mass $M_{\rm max, sub}$ and not on the \emph{host} mass in which the satellite (sub-halo) is orbiting, we could then carry out an identical analysis of the CGPD for satellites where the host halo mass $M$ is replaced by the sub-halo mass $M_{\rm max, sub}$. We would need to add the sub-halo mass function to the distinct halo mass function introduced in Eqn.~\ref{eq:phi}. Then the galaxy property distributions can simply be added,
\begin{equation}
    \phi(\p|z) = \phi(\p|z)_{\rm cen} + \phi(\p|z)_{\rm sat} \,.
\end{equation}
This would add some complexity to the model but would be a straightforward extension. 

\subsubsection{Dependence on higher order halo properties}
Another simplification is that we assume that galaxy properties depend only on halo mass and redshift, which is known to be in contrast with the results of both hydrodynamic simulations and SAMs. Although this ansatz is fundamental to the ubiquitous halo model picture, it has been shown that properties such as galaxy clustering in hydrodynamic simulations depend on halo properties other than mass, an effect commonly referred to as \emph{assembly bias} \citep{Hadzhiyska:2020,Hadzhiyska:2021}. Similarly, it has been shown that some of the dispersion in the stellar mass vs. halo mass relation in simulations is actually due to these higher order halo properties, such as formation history \citep{mitchell2022,Gabrielpillai2021}. 
We are interested in extending the CGPD framework to include additional halo properties, guided by physics-based simulations. 

\subsubsection{Evolution across halo masses and redshifts}

In the main results section, we fit GMM on galaxies for each redshift and halo mass bin independently.
However, a continuous evolution of the galaxy distribution across those discrete bins should be expected.
If the evolution of GMM parameters could be characterized with another parametric model, then the model could be further simplified.
This would be important and possibly necessary for using the approach for statistical inference in the future.

Here we briefly discuss the evolution of galaxy properties across halo masses and redshifts.
To avoid the side effects from possible local maxima of the likelihood function, instead of purely random initialization, we use the best-fit parameter values from the previous bin to initialize the parameters for the next bin.
For GMM with just one component, the best-fit mean vector and covariance matrix are simply the sample mean and covariance of the data.
We notice that the mean of the galaxy properties evolves smoothly across both halo masses and redshifts, while the variances and covariances can change dramatically.
For multiple Gaussian components, an additional challenge is that we need to track each component individually, and there are different approaches for this.
For example, we could sort the components based on their weights, and assume that the component with the highest weight in the current bin always corresponds to the highest-weight one in the next bin, and so on for other components with lower weights.
There are also different definitions of the distance between two Gaussian distributions, based on which we could map one component in the current bin to the closest one in the next bin.
However, this approach may not be very robust since the components could cross each others trajectories.
In our case, we use the best-fit parameters from the current bin as the initial values for the next bin; then the new data can be considered as introducing an update to the current parameters.
From this point of view, the order of parameters remains unchanged across bins.
However, having tried all these approaches, we still found discontinuous behavior of the Gaussian components for not only the covariances but also the mean vectors.
We leave further investigation for future work.

\section{Conclusions} \label{sec:conclusions}

Traditional galaxy surveys and the relatively recent LIM techniques are both powerful methods to study the formation and evolution of galaxies, one of the most fundamental and important problems in astrophysics and cosmology. 
Motivated by the necessity of developing a physically motivated, computationally efficient framework that can describe both galaxy and LIM observations, we propose and characterize a new approach based on the CGPD $\cgpd{\p}$ function, which describes the probability of a halo with mass $M$ at redshift $z$ to host a galaxy with properties $\p$.
When combined with the Line Luminosity Relation (LLR) $L_i(\p,z)$ function developed in a companion work~\citep{Breysse2022}, a highly efficient and flexible pipeline that connects halos with an n-dimensional distribution of galaxy physical properties, and with LIM observables, can be constructed.

As a distribution function in the 5-D galaxy property space that we have chosen, the CGPD encapsulates many important galaxy distribution functions and scaling relations etc. Motivated by the form of the CGPD seen in two separate physics-based cosmological simulations (the Santa Cruz SAM and IllustrisTNG), we discuss a parameterized characterization of CGPD using GMM. We find the following main results: 
\begin{itemize}
    \item With just a few ($\lesssim 3$) components, GMM is capable of reproducing the CGPD distribution from both the SAM and TNG to high accuracy.
    \item The GMM based CGPD yields estimates of common summary statistics of interest including 2-D property relations (scaling relations), property distribution functions, and integrated cosmic densities, with a typical accuracy of a few percent.
    \item For each bin in halo mass and redshift, we find that the 5-D property distribution can be reduced to only 3 dimensions using PCA, with a 10\% loss of information. 
\end{itemize}

The galaxy properties in this work were selected based on their importance for predicting the luminosity of certain emission lines, but a similar approach could be used for other galaxy properties, including galaxy photometry as measured in galaxy surveys. 

The next step is to construct a statistical inference pipeline for LIM observations. For a 5-D Gaussian distribution with a full covariance matrix, we have 20 free parameters including 5 parameters in the mean vector and 15 independent parameters in the symmetric covariance matrix. Then if we use 3 Gaussian components for each halo mass bin and 15 mass bins for each redshift as in our main analysis, we would have 900 parameters for the GMM at each redshift. However, these parameters are expected to be highly degenerate. We have shown that the data dimensionality can be reduced significantly using PCA.
Another possible approach for doing implicit likelihood inference in high dimensional parameter space is the marginal posterior density with moment networks as proposed in~\cite{Jeffrey2020}. 

\section{Acknowledgements}
We thank the IllustrisTNG team for granting the JupyterLab access to the simulations, and Dylan Nelson for help accessing the TNG data.
ARP was supported by NASA under award numbers 80NSSC18K1014, NNH17ZDA001N, and 80NSSC22K0666, and by the NSF under award number 2108411. ARP and rss are supported by the Simons Foundation. 
The numerical computations in this work were performed in part using the NYU Greene High Performance Computing cluster.


\bibliography{cgpd}{}

\begin{thebibliography}{}
\expandafter\ifx\csname natexlab\endcsname\relax\def\natexlab#1{#1}\fi
\providecommand{\url}[1]{\href{#1}{#1}}
\providecommand{\dodoi}[1]{doi:~\href{http://doi.org/#1}{\nolinkurl{#1}}}
\providecommand{\doeprint}[1]{\href{http://ascl.net/#1}{\nolinkurl{http://ascl.net/#1}}}
\providecommand{\doarXiv}[1]{\href{https://arxiv.org/abs/#1}{\nolinkurl{https://arxiv.org/abs/#1}}}

\bibitem[{{Alam} {et~al.}(2017){Alam}, {Ata}, {Bailey}, {Beutler}, {Bizyaev},
  {Blazek}, {Bolton}, {Brownstein}, {Burden}, {Chuang}, {Comparat}, {Cuesta},
  {Dawson}, {Eisenstein}, {Escoffier}, {Gil-Mar{\'\i}n}, {Grieb}, {Hand}, {Ho},
  {Kinemuchi}, {Kirkby}, {Kitaura}, {Malanushenko}, {Malanushenko}, {Maraston},
  {McBride}, {Nichol}, {Olmstead}, {Oravetz}, {Padmanabhan},
  {Palanque-Delabrouille}, {Pan}, {Pellejero-Ibanez}, {Percival}, {Petitjean},
  {Prada}, {Price-Whelan}, {Reid}, {Rodr{\'\i}guez-Torres}, {Roe}, {Ross},
  {Ross}, {Rossi}, {Rubi{\~n}o-Mart{\'\i}n}, {Saito}, {Salazar-Albornoz},
  {Samushia}, {S{\'a}nchez}, {Satpathy}, {Schlegel}, {Schneider},
  {Sc{\'o}ccola}, {Seo}, {Sheldon}, {Simmons}, {Slosar}, {Strauss}, {Swanson},
  {Thomas}, {Tinker}, {Tojeiro}, {Maga{\~n}a}, {Vazquez}, {Verde}, {Wake},
  {Wang}, {Weinberg}, {White}, {Wood-Vasey}, {Y{\`e}che}, {Zehavi}, {Zhai}, \&
  {Zhao}}]{2017MNRAS.470.2617A}
{Alam}, S., {Ata}, M., {Bailey}, S., {et~al.} 2017, \mnras, 470, 2617,
  \dodoi{10.1093/mnras/stx721}

\bibitem[{{Alsing} {et~al.}(2018){Alsing}, {Wandelt}, \& {Feeney}}]{Alsing2018}
{Alsing}, J., {Wandelt}, B., \& {Feeney}, S. 2018, \mnras, 477, 2874,
  \dodoi{10.1093/mnras/sty819}

\bibitem[{{Baldry} {et~al.}(2008){Baldry}, {Glazebrook}, \&
  {Driver}}]{Baldry2008}
{Baldry}, I.~K., {Glazebrook}, K., \& {Driver}, S.~P. 2008, \mnras, 388, 945,
  \dodoi{10.1111/j.1365-2966.2008.13348.x}

\bibitem[{{Behroozi} {et~al.}(2019){Behroozi}, {Wechsler}, {Hearin}, \&
  {Conroy}}]{Behroozi2019}
{Behroozi}, P., {Wechsler}, R.~H., {Hearin}, A.~P., \& {Conroy}, C. 2019,
  \mnras, 488, 3143, \dodoi{10.1093/mnras/stz1182}

\bibitem[{{Bernal} \& {Kovetz}(2022)}]{bernal:2022}
{Bernal}, J.~L., \& {Kovetz}, E.~D. 2022, \aapr, 30, 5,
  \dodoi{10.1007/s00159-022-00143-0}

\bibitem[{{Breysse} {et~al.}(2023)}]{Breysse2022}
{Breysse}, P., {et~al.} 2023, In preparation

\bibitem[{{Calette} {et~al.}(2018){Calette}, {Avila-Reese},
  {Rodr{\'\i}guez-Puebla}, {Hern{\'a}ndez-Toledo}, \&
  {Papastergis}}]{Calette2018}
{Calette}, A.~R., {Avila-Reese}, V., {Rodr{\'\i}guez-Puebla}, A.,
  {Hern{\'a}ndez-Toledo}, H., \& {Papastergis}, E. 2018, \rmxaa, 54, 443,
  \dodoi{10.48550/arXiv.1803.07692}

\bibitem[{{Cranmer} {et~al.}(2020){Cranmer}, {Brehmer}, \&
  {Louppe}}]{2020PNAS..11730055C}
{Cranmer}, K., {Brehmer}, J., \& {Louppe}, G. 2020, Proceedings of the National
  Academy of Science, 117, 30055, \dodoi{10.1073/pnas.1912789117}

\bibitem[{Dempster {et~al.}(1977)Dempster, Laird, \& Rubin}]{Dempster1977}
Dempster, A.~P., Laird, N.~M., \& Rubin, D.~B. 1977, Journal of the Royal
  Statistical Society. Series B (Methodological), 39, 1.
\newblock \url{http://www.jstor.org/stable/2984875}

\bibitem[{{Diemer} {et~al.}(2018){Diemer}, {Stevens}, {Forbes}, {Marinacci},
  {Hernquist}, {Lagos}, {Sternberg}, {Pillepich}, {Nelson}, {Popping},
  {Villaescusa-Navarro}, {Torrey}, \& {Vogelsberger}}]{Diemer2018}
{Diemer}, B., {Stevens}, A. R.~H., {Forbes}, J.~C., {et~al.} 2018, \apjs, 238,
  33, \dodoi{10.3847/1538-4365/aae387}

\bibitem[{{Diemer} {et~al.}(2019){Diemer}, {Stevens}, {Lagos}, {Calette},
  {Tacchella}, {Hernquist}, {Marinacci}, {Nelson}, {Pillepich},
  {Rodriguez-Gomez}, {Villaescusa-Navarro}, \& {Vogelsberger}}]{Diemer2019}
{Diemer}, B., {Stevens}, A. R.~H., {Lagos}, C. d.~P., {et~al.} 2019, \mnras,
  487, 1529, \dodoi{10.1093/mnras/stz1323}

\bibitem[{{Forbes} {et~al.}(2014){Forbes}, {Krumholz}, {Burkert}, \&
  {Dekel}}]{Forbes2014}
{Forbes}, J.~C., {Krumholz}, M.~R., {Burkert}, A., \& {Dekel}, A. 2014, \mnras,
  443, 168, \dodoi{10.1093/mnras/stu1142}

\bibitem[{{Forbes} {et~al.}(2019){Forbes}, {Krumholz}, \&
  {Speagle}}]{2019MNRAS.487.3581F}
{Forbes}, J.~C., {Krumholz}, M.~R., \& {Speagle}, J.~S. 2019, \mnras, 487,
  3581, \dodoi{10.1093/mnras/stz1473}

\bibitem[{{Foreman-Mackey}(2016)}]{Foreman-Mackey2016}
{Foreman-Mackey}, D. 2016, The Journal of Open Source Software, 1, 24,
  \dodoi{10.21105/joss.00024}

\bibitem[{{Gabrielpillai} {et~al.}(2022){Gabrielpillai}, {Somerville}, {Genel},
  {Rodriguez-Gomez}, {Pandya}, {Yung}, \& {Hernquist}}]{Gabrielpillai2021}
{Gabrielpillai}, A., {Somerville}, R.~S., {Genel}, S., {et~al.} 2022, \mnras,
  517, 6091, \dodoi{10.1093/mnras/stac2297}

\bibitem[{{Hadzhiyska} {et~al.}(2020){Hadzhiyska}, {Bose}, {Eisenstein},
  {Hernquist}, \& {Spergel}}]{Hadzhiyska:2020}
{Hadzhiyska}, B., {Bose}, S., {Eisenstein}, D., {Hernquist}, L., \& {Spergel},
  D.~N. 2020, \mnras, 493, 5506, \dodoi{10.1093/mnras/staa623}

\bibitem[{{Hadzhiyska} {et~al.}(2021){Hadzhiyska}, {Liu}, {Somerville},
  {Gabrielpillai}, {Bose}, {Eisenstein}, \& {Hernquist}}]{Hadzhiyska:2021}
{Hadzhiyska}, B., {Liu}, S., {Somerville}, R.~S., {et~al.} 2021, \mnras, 508,
  698, \dodoi{10.1093/mnras/stab2564}

\bibitem[{{Hahn} {et~al.}(2019){Hahn}, {Starkenburg}, {Choi}, {Dav{\'e}},
  {Dickey}, {Geha}, {Genel}, {Hayward}, {Maller}, {Mandyam}, {Pandya},
  {Popping}, {Rafieferantsoa}, {Somerville}, \& {Tinker}}]{2019ApJ...872..160H}
{Hahn}, C., {Starkenburg}, T.~K., {Choi}, E., {et~al.} 2019, \apj, 872, 160,
  \dodoi{10.3847/1538-4357/aafedd}

\bibitem[{{Henriques} {et~al.}(2009){Henriques}, {Thomas}, {Oliver}, \&
  {Roseboom}}]{2009MNRAS.396..535H}
{Henriques}, B. M.~B., {Thomas}, P.~A., {Oliver}, S., \& {Roseboom}, I. 2009,
  \mnras, 396, 535, \dodoi{10.1111/j.1365-2966.2009.14730.x}

\bibitem[{{Henriques} {et~al.}(2013){Henriques}, {White}, {Thomas}, {Angulo},
  {Guo}, {Lemson}, \& {Springel}}]{2013MNRAS.431.3373H}
{Henriques}, B. M.~B., {White}, S. D.~M., {Thomas}, P.~A., {et~al.} 2013,
  \mnras, 431, 3373, \dodoi{10.1093/mnras/stt415}

\bibitem[{{Ivezi{\'c}} {et~al.}(2019){Ivezi{\'c}}, {Connolly}, {Vanderplas}, \&
  {Gray}}]{astroMLText}
{Ivezi{\'c}}, {\v Z}., {Connolly}, A., {Vanderplas}, J., \& {Gray}, A. 2019,
  Statistics, Data Mining and Machine Learning in Astronomy (Princeton
  University Press)

\bibitem[{{Jeffrey} \& {Wandelt}(2020)}]{Jeffrey2020}
{Jeffrey}, N., \& {Wandelt}, B.~D. 2020, arXiv e-prints, arXiv:2011.05991.
\newblock \doarXiv{2011.05991}

\bibitem[{{Koekemoer} {et~al.}(2011){Koekemoer}, {Faber}, {Ferguson}, {Grogin},
  {Kocevski}, {Koo}, {Lai}, {Lotz}, {Lucas}, {McGrath}, {Ogaz}, {Rajan},
  {Riess}, {Rodney}, {Strolger}, {Casertano}, {Castellano}, {Dahlen},
  {Dickinson}, {Dolch}, {Fontana}, {Giavalisco}, {Grazian}, {Guo}, {Hathi},
  {Huang}, {van der Wel}, {Yan}, {Acquaviva}, {Alexander}, {Almaini}, {Ashby},
  {Barden}, {Bell}, {Bournaud}, {Brown}, {Caputi}, {Cassata}, {Challis},
  {Chary}, {Cheung}, {Cirasuolo}, {Conselice}, {Roshan Cooray}, {Croton},
  {Daddi}, {Dav{\'e}}, {de Mello}, {de Ravel}, {Dekel}, {Donley}, {Dunlop},
  {Dutton}, {Elbaz}, {Fazio}, {Filippenko}, {Finkelstein}, {Frazer}, {Gardner},
  {Garnavich}, {Gawiser}, {Gruetzbauch}, {Hartley}, {H{\"a}ussler},
  {Herrington}, {Hopkins}, {Huang}, {Jha}, {Johnson}, {Kartaltepe},
  {Khostovan}, {Kirshner}, {Lani}, {Lee}, {Li}, {Madau}, {McCarthy},
  {McIntosh}, {McLure}, {McPartland}, {Mobasher}, {Moreira}, {Mortlock},
  {Moustakas}, {Mozena}, {Nandra}, {Newman}, {Nielsen}, {Niemi}, {Noeske},
  {Papovich}, {Pentericci}, {Pope}, {Primack}, {Ravindranath}, {Reddy},
  {Renzini}, {Rix}, {Robaina}, {Rosario}, {Rosati}, {Salimbeni}, {Scarlata},
  {Siana}, {Simard}, {Smidt}, {Snyder}, {Somerville}, {Spinrad}, {Straughn},
  {Telford}, {Teplitz}, {Trump}, {Vargas}, {Villforth}, {Wagner}, {Wandro},
  {Wechsler}, {Weiner}, {Wiklind}, {Wild}, {Wilson}, {Wuyts}, \&
  {Yun}}]{Koekemoer2011}
{Koekemoer}, A.~M., {Faber}, S.~M., {Ferguson}, H.~C., {et~al.} 2011, \apjs,
  197, 36, \dodoi{10.1088/0067-0049/197/2/36}

\bibitem[{{Kovetz} {et~al.}(2017){Kovetz}, {Viero}, {Lidz}, {Newburgh},
  {Rahman}, {Switzer}, {Kamionkowski}, {Aguirre}, {Alvarez}, {Bock}, {Bond},
  {Bower}, {Bradford}, {Breysse}, {Bull}, {Chang}, {Cheng}, {Chung}, {Cleary},
  {Corray}, {Crites}, {Croft}, {Dor{\'e}}, {Eastwood}, {Ferrara}, {Fonseca},
  {Jacobs}, {Keating}, {Lagache}, {Lakhlani}, {Liu}, {Moodley}, {Murray},
  {P{\'e}nin}, {Popping}, {Pullen}, {Reichers}, {Saito}, {Saliwanchik},
  {Santos}, {Somerville}, {Stacey}, {Stein}, {Villaescusa-Navarro}, {Visbal},
  {Weltman}, {Wolz}, \& {Zemcov}}]{Kovetz2017}
{Kovetz}, E.~D., {Viero}, M.~P., {Lidz}, A., {et~al.} 2017, arXiv e-prints,
  arXiv:1709.09066.
\newblock \doarXiv{1709.09066}

\bibitem[{{Krumholz}(2014)}]{despotic}
{Krumholz}, M.~R. 2014, \mnras, 437, 1662, \dodoi{10.1093/mnras/stt2000}

\bibitem[{{Lange} {et~al.}(2015){Lange}, {Driver}, {Robotham}, {Kelvin},
  {Graham}, {Alpaslan}, {Andrews}, {Baldry}, {Bamford}, {Bland-Hawthorn},
  {Brough}, {Cluver}, {Conselice}, {Davies}, {Haeussler}, {Konstantopoulos},
  {Loveday}, {Moffett}, {Norberg}, {Phillipps}, {Taylor},
  {L{\'o}pez-S{\'a}nchez}, \& {Wilkins}}]{Lange2015}
{Lange}, R., {Driver}, S.~P., {Robotham}, A. S.~G., {et~al.} 2015, \mnras, 447,
  2603, \dodoi{10.1093/mnras/stu2467}

\bibitem[{{Lara-L{\'o}pez} {et~al.}(2010){Lara-L{\'o}pez}, {Cepa},
  {Bongiovanni}, {P{\'e}rez Garc{\'\i}a}, {Ederoclite}, {Casta{\~n}eda},
  {Fern{\'a}ndez Lorenzo}, {Povi{\'c}}, \&
  {S{\'a}nchez-Portal}}]{Lara-Lopez2010}
{Lara-L{\'o}pez}, M.~A., {Cepa}, J., {Bongiovanni}, A., {et~al.} 2010, \aap,
  521, L53, \dodoi{10.1051/0004-6361/201014803}

\bibitem[{{Leja} {et~al.}(2022){Leja}, {Speagle}, {Ting}, {Johnson}, {Conroy},
  {Whitaker}, {Nelson}, {Dokkum}, \& {Franx}}]{Leja2022}
{Leja}, J., {Speagle}, J.~S., {Ting}, Y.-S., {et~al.} 2022, \apj, 936, 165,
  \dodoi{10.3847/1538-4357/ac887d}

\bibitem[{{Leung} {et~al.}(2020){Leung}, {Olsen}, {Somerville}, {Dav{\'e}},
  {Greve}, {Hayward}, {Narayanan}, \& {Popping}}]{Leung}
{Leung}, T.~K.~D., {Olsen}, K.~P., {Somerville}, R.~S., {et~al.} 2020, \apj,
  905, 102, \dodoi{10.3847/1538-4357/abc25e}

\bibitem[{{Li} {et~al.}(2016){Li}, {Wechsler}, {Devaraj}, \& {Church}}]{Li2016}
{Li}, T.~Y., {Wechsler}, R.~H., {Devaraj}, K., \& {Church}, S.~E. 2016, \apj,
  817, 169, \dodoi{10.3847/0004-637X/817/2/169}

\bibitem[{{Lu} {et~al.}(2012){Lu}, {Mo}, {Katz}, \&
  {Weinberg}}]{2012MNRAS.421.1779L}
{Lu}, Y., {Mo}, H.~J., {Katz}, N., \& {Weinberg}, M.~D. 2012, \mnras, 421,
  1779, \dodoi{10.1111/j.1365-2966.2012.20435.x}

\bibitem[{{Lu} {et~al.}(2014){Lu}, {Mo}, {Lu}, {Katz}, \&
  {Weinberg}}]{2014MNRAS.443.1252L}
{Lu}, Y., {Mo}, H.~J., {Lu}, Z., {Katz}, N., \& {Weinberg}, M.~D. 2014, \mnras,
  443, 1252, \dodoi{10.1093/mnras/stu1200}

\bibitem[{{Lu} {et~al.}(2011){Lu}, {Mo}, {Weinberg}, \&
  {Katz}}]{2011MNRAS.416.1949L}
{Lu}, Y., {Mo}, H.~J., {Weinberg}, M.~D., \& {Katz}, N. 2011, \mnras, 416,
  1949, \dodoi{10.1111/j.1365-2966.2011.19170.x}

\bibitem[{{Madau} \& {Dickinson}(2014)}]{Madau2014}
{Madau}, P., \& {Dickinson}, M. 2014, \araa, 52, 415,
  \dodoi{10.1146/annurev-astro-081811-125615}

\bibitem[{{Marinacci} {et~al.}(2018){Marinacci}, {Vogelsberger}, {Pakmor},
  {Torrey}, {Springel}, {Hernquist}, {Nelson}, {Weinberger}, {Pillepich},
  {Naiman}, \& {Genel}}]{TNG-3}
{Marinacci}, F., {Vogelsberger}, M., {Pakmor}, R., {et~al.} 2018, \mnras, 480,
  5113, \dodoi{10.1093/mnras/sty2206}

\bibitem[{{Mitchell} \& {Schaye}(2022)}]{mitchell2022}
{Mitchell}, P.~D., \& {Schaye}, J. 2022, \mnras, 511, 2948,
  \dodoi{10.1093/mnras/stab3339}

\bibitem[{{Mowla} {et~al.}(2019){Mowla}, {van der Wel}, {van Dokkum}, \&
  {Miller}}]{Mowla2019}
{Mowla}, L., {van der Wel}, A., {van Dokkum}, P., \& {Miller}, T.~B. 2019,
  \apjl, 872, L13, \dodoi{10.3847/2041-8213/ab0379}

\bibitem[{{Murray} {et~al.}(2013){Murray}, {Power}, \& {Robotham}}]{Murray2013}
{Murray}, S.~G., {Power}, C., \& {Robotham}, A.~S.~G. 2013, Astronomy and
  Computing, 3, 23, \dodoi{10.1016/j.ascom.2013.11.001}

\bibitem[{{Naab} \& {Ostriker}(2017)}]{naab2017}
{Naab}, T., \& {Ostriker}, J.~P. 2017, \araa, 55, 59,
  \dodoi{10.1146/annurev-astro-081913-040019}

\bibitem[{{Naiman} {et~al.}(2018){Naiman}, {Pillepich}, {Springel},
  {Ramirez-Ruiz}, {Torrey}, {Vogelsberger}, {Pakmor}, {Nelson}, {Marinacci},
  {Hernquist}, {Weinberger}, \& {Genel}}]{TNG-5}
{Naiman}, J.~P., {Pillepich}, A., {Springel}, V., {et~al.} 2018, \mnras, 477,
  1206, \dodoi{10.1093/mnras/sty618}

\bibitem[{{Nelson} {et~al.}(2018){Nelson}, {Pillepich}, {Springel},
  {Weinberger}, {Hernquist}, {Pakmor}, {Genel}, {Torrey}, {Vogelsberger},
  {Kauffmann}, {Marinacci}, \& {Naiman}}]{TNG-4}
{Nelson}, D., {Pillepich}, A., {Springel}, V., {et~al.} 2018, \mnras, 475, 624,
  \dodoi{10.1093/mnras/stx3040}

\bibitem[{{Noeske} {et~al.}(2007){Noeske}, {Weiner}, {Faber}, {Papovich},
  {Koo}, {Somerville}, {Bundy}, {Conselice}, {Newman}, {Schiminovich}, {Le
  Floc'h}, {Coil}, {Rieke}, {Lotz}, {Primack}, {Barmby}, {Cooper}, {Davis},
  {Ellis}, {Fazio}, {Guhathakurta}, {Huang}, {Kassin}, {Martin}, {Phillips},
  {Rich}, {Small}, {Willmer}, \& {Wilson}}]{Noeske2007}
{Noeske}, K.~G., {Weiner}, B.~J., {Faber}, S.~M., {et~al.} 2007, \apjl, 660,
  L43, \dodoi{10.1086/517926}

\bibitem[{{Olsen} {et~al.}(2021){Olsen}, {Burkhart}, {Mac Low}, {Tre{\ss}},
  {Greve}, {Vizgan}, {Motka}, {Borrow}, {Popping}, {Dav{\'e}}, {Smith}, \&
  {Narayanan}}]{Olsen}
{Olsen}, K.~P., {Burkhart}, B., {Mac Low}, M.-M., {et~al.} 2021, \apj, 922, 88,
  \dodoi{10.3847/1538-4357/ac20d4}

\bibitem[{{Padmanabhan}(2018)}]{Padmanabhan2018}
{Padmanabhan}, H. 2018, \mnras, 475, 1477, \dodoi{10.1093/mnras/stx3250}

\bibitem[{{Padmanabhan}(2019)}]{Padmanabhan2019}
---. 2019, \mnras, 488, 3014, \dodoi{10.1093/mnras/stz1878}

\bibitem[{{Pallottini} {et~al.}(2022){Pallottini}, {Ferrara}, {Gallerani},
  {Behrens}, {Kohandel}, {Carniani}, {Vallini}, {Salvadori}, {Gelli},
  {Sommovigo}, {D'Odorico}, {Di Mascia}, \& {Pizzati}}]{pallottini}
{Pallottini}, A., {Ferrara}, A., {Gallerani}, S., {et~al.} 2022, \mnras, 513,
  5621, \dodoi{10.1093/mnras/stac1281}

\bibitem[{{Pillepich} {et~al.}(2018){Pillepich}, {Nelson}, {Hernquist},
  {Springel}, {Pakmor}, {Torrey}, {Weinberger}, {Genel}, {Naiman}, {Marinacci},
  \& {Vogelsberger}}]{TNG-2}
{Pillepich}, A., {Nelson}, D., {Hernquist}, L., {et~al.} 2018, \mnras, 475,
  648, \dodoi{10.1093/mnras/stx3112}

\bibitem[{{Planck Collaboration} {et~al.}(2016){Planck Collaboration}, {Ade},
  {Aghanim}, {Arnaud}, {Ashdown}, {Aumont}, {Baccigalupi}, {Banday},
  {Barreiro}, {Bartlett}, {Bartolo}, {Battaner}, {Battye}, {Benabed},
  {Beno{\^\i}t}, {Benoit-L{\'e}vy}, {Bernard}, {Bersanelli}, {Bielewicz},
  {Bock}, {Bonaldi}, {Bonavera}, {Bond}, {Borrill}, {Bouchet}, {Boulanger},
  {Bucher}, {Burigana}, {Butler}, {Calabrese}, {Cardoso}, {Catalano},
  {Challinor}, {Chamballu}, {Chary}, {Chiang}, {Chluba}, {Christensen},
  {Church}, {Clements}, {Colombi}, {Colombo}, {Combet}, {Coulais}, {Crill},
  {Curto}, {Cuttaia}, {Danese}, {Davies}, {Davis}, {de Bernardis}, {de Rosa},
  {de Zotti}, {Delabrouille}, {D{\'e}sert}, {Di Valentino}, {Dickinson},
  {Diego}, {Dolag}, {Dole}, {Donzelli}, {Dor{\'e}}, {Douspis}, {Ducout},
  {Dunkley}, {Dupac}, {Efstathiou}, {Elsner}, {En{\ss}lin}, {Eriksen},
  {Farhang}, {Fergusson}, {Finelli}, {Forni}, {Frailis}, {Fraisse},
  {Franceschi}, {Frejsel}, {Galeotta}, {Galli}, {Ganga}, {Gauthier}, {Gerbino},
  {Ghosh}, {Giard}, {Giraud-H{\'e}raud}, {Giusarma}, {Gjerl{\o}w},
  {Gonz{\'a}lez-Nuevo}, {G{\'o}rski}, {Gratton}, {Gregorio}, {Gruppuso},
  {Gudmundsson}, {Hamann}, {Hansen}, {Hanson}, {Harrison}, {Helou},
  {Henrot-Versill{\'e}}, {Hern{\'a}ndez-Monteagudo}, {Herranz}, {Hildebrandt},
  {Hivon}, {Hobson}, {Holmes}, {Hornstrup}, {Hovest}, {Huang}, {Huffenberger},
  {Hurier}, {Jaffe}, {Jaffe}, {Jones}, {Juvela}, {Keih{\"a}nen}, {Keskitalo},
  {Kisner}, {Kneissl}, {Knoche}, {Knox}, {Kunz}, {Kurki-Suonio}, {Lagache},
  {L{\"a}hteenm{\"a}ki}, {Lamarre}, {Lasenby}, {Lattanzi}, {Lawrence}, {Leahy},
  {Leonardi}, {Lesgourgues}, {Levrier}, {Lewis}, {Liguori}, {Lilje},
  {Linden-V{\o}rnle}, {L{\'o}pez-Caniego}, {Lubin}, {Mac{\'\i}as-P{\'e}rez},
  {Maggio}, {Maino}, {Mandolesi}, {Mangilli}, {Marchini}, {Maris}, {Martin},
  {Martinelli}, {Mart{\'\i}nez-Gonz{\'a}lez}, {Masi}, {Matarrese}, {McGehee},
  {Meinhold}, {Melchiorri}, {Melin}, {Mendes}, {Mennella}, {Migliaccio},
  {Millea}, {Mitra}, {Miville-Desch{\^e}nes}, {Moneti}, {Montier}, {Morgante},
  {Mortlock}, {Moss}, {Munshi}, {Murphy}, {Naselsky}, {Nati}, {Natoli},
  {Netterfield}, {N{\o}rgaard-Nielsen}, {Noviello}, {Novikov}, {Novikov},
  {Oxborrow}, {Paci}, {Pagano}, {Pajot}, {Paladini}, {Paoletti}, {Partridge},
  {Pasian}, {Patanchon}, {Pearson}, {Perdereau}, {Perotto}, {Perrotta},
  {Pettorino}, {Piacentini}, {Piat}, {Pierpaoli}, {Pietrobon}, {Plaszczynski},
  {Pointecouteau}, {Polenta}, {Popa}, {Pratt}, {Pr{\'e}zeau}, {Prunet},
  {Puget}, {Rachen}, {Reach}, {Rebolo}, {Reinecke}, {Remazeilles}, {Renault},
  {Renzi}, {Ristorcelli}, {Rocha}, {Rosset}, {Rossetti}, {Roudier},
  {Rouill{\'e} d'Orfeuil}, {Rowan-Robinson}, {Rubi{\~n}o-Mart{\'\i}n},
  {Rusholme}, {Said}, {Salvatelli}, {Salvati}, {Sandri}, {Santos},
  {Savelainen}, {Savini}, {Scott}, {Seiffert}, {Serra}, {Shellard}, {Spencer},
  {Spinelli}, {Stolyarov}, {Stompor}, {Sudiwala}, {Sunyaev}, {Sutton},
  {Suur-Uski}, {Sygnet}, {Tauber}, {Terenzi}, {Toffolatti}, {Tomasi},
  {Tristram}, {Trombetti}, {Tucci}, {Tuovinen}, {T{\"u}rler}, {Umana},
  {Valenziano}, {Valiviita}, {Van Tent}, {Vielva}, {Villa}, {Wade}, {Wandelt},
  {Wehus}, {White}, {White}, {Wilkinson}, {Yvon}, {Zacchei}, \&
  {Zonca}}]{Planck2015-cosmo}
{Planck Collaboration}, {Ade}, P.~A.~R., {Aghanim}, N., {et~al.} 2016, \aap,
  594, A13, \dodoi{10.1051/0004-6361/201525830}

\bibitem[{{Popping} {et~al.}(2019){Popping}, {Narayanan}, {Somerville},
  {Faisst}, \& {Krumholz}}]{popping2019}
{Popping}, G., {Narayanan}, D., {Somerville}, R.~S., {Faisst}, A.~L., \&
  {Krumholz}, M.~R. 2019, \mnras, 482, 4906, \dodoi{10.1093/mnras/sty2969}

\bibitem[{{Popping} {et~al.}(2014){Popping}, {Somerville}, \&
  {Trager}}]{Popping2014}
{Popping}, G., {Somerville}, R.~S., \& {Trager}, S.~C. 2014, \mnras, 442, 2398,
  \dodoi{10.1093/mnras/stu991}

\bibitem[{{Porter} {et~al.}(2014){Porter}, {Somerville}, {Primack}, \&
  {Johansson}}]{Porter2014}
{Porter}, L.~A., {Somerville}, R.~S., {Primack}, J.~R., \& {Johansson}, P.~H.
  2014, \mnras, 444, 942, \dodoi{10.1093/mnras/stu1434}

\bibitem[{{Schaan} \& {White}(2021{\natexlab{a}})}]{Schaan2021a}
{Schaan}, E., \& {White}, M. 2021{\natexlab{a}}, \jcap, 2021, 067,
  \dodoi{10.1088/1475-7516/2021/05/067}

\bibitem[{{Schaan} \& {White}(2021{\natexlab{b}})}]{Schaan2021b}
---. 2021{\natexlab{b}}, \jcap, 2021, 068,
  \dodoi{10.1088/1475-7516/2021/05/068}

\bibitem[{{Silva} {et~al.}(2015){Silva}, {Santos}, {Cooray}, \&
  {Gong}}]{Silva2015}
{Silva}, M., {Santos}, M.~G., {Cooray}, A., \& {Gong}, Y. 2015, \apj, 806, 209,
  \dodoi{10.1088/0004-637X/806/2/209}

\bibitem[{{Smit} {et~al.}(2012){Smit}, {Bouwens}, {Franx}, {Illingworth},
  {Labb{\'e}}, {Oesch}, \& {van Dokkum}}]{Smit2012}
{Smit}, R., {Bouwens}, R.~J., {Franx}, M., {et~al.} 2012, \apj, 756, 14,
  \dodoi{10.1088/0004-637X/756/1/14}

\bibitem[{{Somerville} \& {Dav{\'e}}(2015{\natexlab{a}})}]{somerville-dave}
{Somerville}, R.~S., \& {Dav{\'e}}, R. 2015{\natexlab{a}}, \araa, 53, 51,
  \dodoi{10.1146/annurev-astro-082812-140951}

\bibitem[{{Somerville} \& {Dav{\'e}}(2015{\natexlab{b}})}]{Somerville2015b}
---. 2015{\natexlab{b}}, \araa, 53, 51,
  \dodoi{10.1146/annurev-astro-082812-140951}

\bibitem[{{Somerville} {et~al.}(2012){Somerville}, {Gilmore}, {Primack}, \&
  {Dom{\'\i}nguez}}]{Somerville2012}
{Somerville}, R.~S., {Gilmore}, R.~C., {Primack}, J.~R., \& {Dom{\'\i}nguez},
  A. 2012, \mnras, 423, 1992, \dodoi{10.1111/j.1365-2966.2012.20490.x}

\bibitem[{{Somerville} {et~al.}(2008){Somerville}, {Hopkins}, {Cox},
  {Robertson}, \& {Hernquist}}]{Somerville2008}
{Somerville}, R.~S., {Hopkins}, P.~F., {Cox}, T.~J., {Robertson}, B.~E., \&
  {Hernquist}, L. 2008, \mnras, 391, 481,
  \dodoi{10.1111/j.1365-2966.2008.13805.x}

\bibitem[{{Somerville} \& {Kolatt}(1999)}]{Somerville1999a}
{Somerville}, R.~S., \& {Kolatt}, T.~S. 1999, \mnras, 305, 1,
  \dodoi{10.1046/j.1365-8711.1999.02154.x}

\bibitem[{{Somerville} {et~al.}(2015){Somerville}, {Popping}, \&
  {Trager}}]{Somerville2015a}
{Somerville}, R.~S., {Popping}, G., \& {Trager}, S.~C. 2015, \mnras, 453, 4337,
  \dodoi{10.1093/mnras/stv1877}

\bibitem[{{Somerville} \& {Primack}(1999)}]{Somerville1999}
{Somerville}, R.~S., \& {Primack}, J.~R. 1999, \mnras, 310, 1087,
  \dodoi{10.1046/j.1365-8711.1999.03032.x}

\bibitem[{{Somerville} {et~al.}(2021){Somerville}, {Olsen}, {Yung}, {Pacifici},
  {Ferguson}, {Behroozi}, {Osborne}, {Wechsler}, {Pandya}, {Faber}, {Primack},
  \& {Dekel}}]{Somerville2021}
{Somerville}, R.~S., {Olsen}, C., {Yung}, L.~Y.~A., {et~al.} 2021, \mnras, 502,
  4858, \dodoi{10.1093/mnras/stab231}

\bibitem[{{Springel} {et~al.}(2018){Springel}, {Pakmor}, {Pillepich},
  {Weinberger}, {Nelson}, {Hernquist}, {Vogelsberger}, {Genel}, {Torrey},
  {Marinacci}, \& {Naiman}}]{TNG-1}
{Springel}, V., {Pakmor}, R., {Pillepich}, A., {et~al.} 2018, \mnras, 475, 676,
  \dodoi{10.1093/mnras/stx3304}

\bibitem[{{Stevens} {et~al.}(2021){Stevens}, {Lagos}, {Cortese}, {Catinella},
  {Diemer}, {Nelson}, {Pillepich}, {Hernquist}, {Marinacci}, \&
  {Vogelsberger}}]{Stevens2021}
{Stevens}, A. R.~H., {Lagos}, C. d.~P., {Cortese}, L., {et~al.} 2021, \mnras,
  502, 3158, \dodoi{10.1093/mnras/staa3662}

\bibitem[{{Sun} {et~al.}(2019){Sun}, {Hensley}, {Chang}, {Dor{\'e}}, \&
  {Serra}}]{Sun2019}
{Sun}, G., {Hensley}, B.~S., {Chang}, T.-C., {Dor{\'e}}, O., \& {Serra}, P.
  2019, \apj, 887, 142, \dodoi{10.3847/1538-4357/ab55df}

\bibitem[{{Tinker} {et~al.}(2008){Tinker}, {Kravtsov}, {Klypin}, {Abazajian},
  {Warren}, {Yepes}, {Gottl{\"o}ber}, \& {Holz}}]{Tinker2008}
{Tinker}, J., {Kravtsov}, A.~V., {Klypin}, A., {et~al.} 2008, \apj, 688, 709,
  \dodoi{10.1086/591439}

\bibitem[{{Tremonti} {et~al.}(2004){Tremonti}, {Heckman}, {Kauffmann},
  {Brinchmann}, {Charlot}, {White}, {Seibert}, {Peng}, {Schlegel}, {Uomoto},
  {Fukugita}, \& {Brinkmann}}]{Tremonti2004}
{Tremonti}, C.~A., {Heckman}, T.~M., {Kauffmann}, G., {et~al.} 2004, \apj, 613,
  898, \dodoi{10.1086/423264}

\bibitem[{{Wechsler} \& {Tinker}(2018)}]{wechsler-tinker:2018}
{Wechsler}, R.~H., \& {Tinker}, J.~L. 2018, \araa, 56, 435,
  \dodoi{10.1146/annurev-astro-081817-051756}

\bibitem[{{Yang} {et~al.}(2022){Yang}, {Popping}, {Somerville}, {Pullen},
  {Breysse}, \& {Maniyar}}]{yang2022}
{Yang}, S., {Popping}, G., {Somerville}, R.~S., {et~al.} 2022, \apj, 929, 140,
  \dodoi{10.3847/1538-4357/ac5d57}

\bibitem[{{Yang} {et~al.}(2021){Yang}, {Somerville}, {Pullen}, {Popping},
  {Breysse}, \& {Maniyar}}]{Yang2021}
{Yang}, S., {Somerville}, R.~S., {Pullen}, A.~R., {et~al.} 2021, \apj, 911,
  132, \dodoi{10.3847/1538-4357/abec75}

\bibitem[{{York} {et~al.}(2000){York}, {Adelman}, {Anderson}, {Anderson},
  {Annis}, {Bahcall}, {Bakken}, {Barkhouser}, {Bastian}, {Berman}, {Boroski},
  {Bracker}, {Briegel}, {Briggs}, {Brinkmann}, {Brunner}, {Burles}, {Carey},
  {Carr}, {Castander}, {Chen}, {Colestock}, {Connolly}, {Crocker}, {Csabai},
  {Czarapata}, {Davis}, {Doi}, {Dombeck}, {Eisenstein}, {Ellman}, {Elms},
  {Evans}, {Fan}, {Federwitz}, {Fiscelli}, {Friedman}, {Frieman}, {Fukugita},
  {Gillespie}, {Gunn}, {Gurbani}, {de Haas}, {Haldeman}, {Harris}, {Hayes},
  {Heckman}, {Hennessy}, {Hindsley}, {Holm}, {Holmgren}, {Huang}, {Hull},
  {Husby}, {Ichikawa}, {Ichikawa}, {Ivezi{\'c}}, {Kent}, {Kim}, {Kinney},
  {Klaene}, {Kleinman}, {Kleinman}, {Knapp}, {Korienek}, {Kron}, {Kunszt},
  {Lamb}, {Lee}, {Leger}, {Limmongkol}, {Lindenmeyer}, {Long}, {Loomis},
  {Loveday}, {Lucinio}, {Lupton}, {MacKinnon}, {Mannery}, {Mantsch}, {Margon},
  {McGehee}, {McKay}, {Meiksin}, {Merelli}, {Monet}, {Munn}, {Narayanan},
  {Nash}, {Neilsen}, {Neswold}, {Newberg}, {Nichol}, {Nicinski}, {Nonino},
  {Okada}, {Okamura}, {Ostriker}, {Owen}, {Pauls}, {Peoples}, {Peterson},
  {Petravick}, {Pier}, {Pope}, {Pordes}, {Prosapio}, {Rechenmacher}, {Quinn},
  {Richards}, {Richmond}, {Rivetta}, {Rockosi}, {Ruthmansdorfer}, {Sandford},
  {Schlegel}, {Schneider}, {Sekiguchi}, {Sergey}, {Shimasaku}, {Siegmund},
  {Smee}, {Smith}, {Snedden}, {Stone}, {Stoughton}, {Strauss}, {Stubbs},
  {SubbaRao}, {Szalay}, {Szapudi}, {Szokoly}, {Thakar}, {Tremonti}, {Tucker},
  {Uomoto}, {Vanden Berk}, {Vogeley}, {Waddell}, {Wang}, {Watanabe},
  {Weinberg}, {Yanny}, {Yasuda}, \& {SDSS Collaboration}}]{York2000}
{York}, D.~G., {Adelman}, J., {Anderson}, John~E., J., {et~al.} 2000, \aj, 120,
  1579, \dodoi{10.1086/301513}

\bibitem[{{Yung} {et~al.}(2019{\natexlab{a}}){Yung}, {Somerville},
  {Finkelstein}, {Popping}, \& {Dav{\'e}}}]{Yung2019a}
{Yung}, L.~Y.~A., {Somerville}, R.~S., {Finkelstein}, S.~L., {Popping}, G., \&
  {Dav{\'e}}, R. 2019{\natexlab{a}}, \mnras, 483, 2983,
  \dodoi{10.1093/mnras/sty3241}

\bibitem[{{Yung} {et~al.}(2019{\natexlab{b}}){Yung}, {Somerville}, {Popping},
  {Finkelstein}, {Ferguson}, \& {Dav{\'e}}}]{Yung2019b}
{Yung}, L.~Y.~A., {Somerville}, R.~S., {Popping}, G., {et~al.}
  2019{\natexlab{b}}, \mnras, 490, 2855, \dodoi{10.1093/mnras/stz2755}

\end{thebibliography}
\bibliographystyle{aasjournal}

\end{document}